\documentclass[prd,amsmath,amssymb,longbibliography,12pt,nofootinbib]{revtex4-2}
\usepackage[normalem]{ulem}
\usepackage{soul}
\usepackage[utf8]{inputenc}
\usepackage{lineno,hyperref}
\usepackage{amsfonts,amssymb,amsmath}
\usepackage{epsfig}
\usepackage{mathrsfs}
\usepackage{amssymb}
\usepackage{graphicx}
\usepackage{amsmath}

\usepackage{xcolor}
\usepackage{color}
\usepackage[normalem]{ulem}
\usepackage{soul}
\usepackage{appendix}

\usepackage{xcolor}
\begin{document}

\title{Krylov complexity and oscillatory quantum dynamics in two-field inflation}
\author{Tao Li$^{1}$}
\email{2022700450@stu.jsu.edu.cn}
\author{Lei-Hua Liu$^{2}$}
\email{ liuleihua8899@hotmail.com}
\author{Hai-Bing Fu$^{1}$}

\affiliation{$^{1}$Department of Physics, Guizhou Minzu University, Guiyang 550025, China}
\affiliation{$^{2}$Department of Physics, College of Physics, Mechanical and Electrical Engineering, Jishou University, Jishou 416000, China}

\begin{abstract}
 
We investigate Krylov-space dynamics of primordial perturbations in two-field inflation. Starting from the quadratic action for coupled curvature and entropy perturbations, we solve the mode equations with Bunch–Davies initial conditions and construct the corresponding quadratic Hamiltonian. The Gaussian quantum state is described by a four-mode squeezed state with two squeezing channels. We derive the evolution of the squeezing strength and squeezing angle and use them to evaluate the Krylov complexity, Krylov entropy, and channel-resolved Lanczos coefficients. Our numerical results show that the two channels remain weakly squeezed inside the horizon but develop different squeezing histories after horizon exit. The second channel develops strong squeezing earlier and dominates the intermediate evolution, whereas the first channel shows a delayed enhancement. The corresponding Krylov complexities exhibit distinct oscillatory structures across horizon crossing and pronounced fluctuations in the strongly squeezed superhorizon regime. These features arise from the phase-sensitive squeezing dynamics and reflect the different responses of the two channels to curvature–entropy coupling.

\end{abstract}

\maketitle{}

\section{Introduction}
\label{introduction}

Inflation provides a simple mechanism for generating the primordial fluctuations that seed the large-scale structure of the Universe. However, the quantum-information properties of these fluctuations are still not fully understood. In recent years, quantum complexity has provided a new way to study quantum dynamics. Krylov complexity is particularly useful because it describes the spreading of an initially simple operator or quantum state in Krylov space. The modern formulation is based on the Lanczos algorithm. The universal operator growth hypothesis was proposed to relate the growth of the Lanczos coefficients to quantum chaos \cite{Parker:2018yvk}. This framework was later developed in quantum many-body systems, conformal field theories, quantum field theory, open systems, and holography \cite{Barbon:2019wsy,Dymarsky:2019elm,Mohan:2023btr,Avdoshkin:2019trj,Dymarsky:2021bjq,Caputa:2021sib,Rabinovici:2021qqt,Muck:2022xfc,Liu:2022god,Avdoshkin:2022xuw}. In particular, the growth of the Lanczos coefficients provides a direct measure of operator spreading, while Krylov complexity describes how far the quantum dynamics spreads away from its initial state \cite{Parker:2018yvk,Barbon:2019wsy,Dymarsky:2019elm,Avdoshkin:2019trj,Dymarsky:2021bjq,Caputa:2021sib,Rabinovici:2021qqt,Muck:2022xfc,Liu:2022god,Avdoshkin:2022xuw}.

The application of Krylov complexity to cosmology is much more recent. Krylov complexity was subsequently applied to the inflationary patch of de Sitter space using the two-mode squeezed-state description \cite{Adhikari:2022oxr,Adhikari:2022whf}. These studies showed that cosmological squeezing can be mapped into Krylov-space evolution and established a direct relation between Krylov complexity and the mean particle number. Further studies investigated the Krylov complexity and related quantum complexity measures of primordial perturbations during inflation, including closed- and open-system descriptions, modified dispersion relations, thermal effects, K-essence models, and non-trivial sound speed \cite{Li:2024kfm,Li:2024ljz,Zhai:2024odw,Liu:2025caj,Li:2026jxx,Li:2023ekd,Li:2024iji,Liu:2021nzx,Li:2021kfq}. Effective gravitational descriptions of inflation motivated by quantum corrections have also been studied in related contexts \cite{Liu:2018hno}. In particular, oscillatory complexity around horizon crossing and the dependence of the complexity growth on squeezing dynamics were found to provide useful diagnostics of inflationary evolution \cite{Li:2024kfm}. These developments were further extended to different stages of the early Universe and to open-system descriptions of Krylov complexity and entropy \cite{Zhai:2024odw,Li:2024ljz,Liu:2025caj}.

Most cosmological studies of Krylov complexity have nevertheless considered a single scalar degree of freedom or an effectively single-field description. The corresponding quantum state is usually described by a two-mode squeezed state associated with the momentum pair \(({k},-{k})\). This description is very useful, but it cannot describe quantum information transfer between different scalar degrees of freedom. The next natural step is therefore to study Krylov complexity in multi-field inflation.

Multi-field inflation has a long history. Early studies considered multi-component and double-inflation scenarios \cite{Starobinsky:1985ibc,Silk:1986vc,Garcia-Bellido:1995hsq,Liddle:1998jc}. The perturbation theory of these models was subsequently developed in detail \cite{Sasaki:1998ug,Langlois:1999dw,GarciadeAndrade:2014czy,Wands:2002bn,GrootNibbelink:2001qt,Wands:2007bd}. An important result is the separation of perturbations into adiabatic and entropy directions. In two-field inflation, the adiabatic mode describes fluctuations along the background trajectory, while the entropy mode describes fluctuations orthogonal to it \cite{Wands:2002bn,GrootNibbelink:2001qt,Wands:2007bd}. A curved background trajectory allows the entropy perturbation to source the curvature perturbation even after horizon exit \cite{Wands:2007bd,Vernizzi:2006ve,Lyth:2005fi}. Therefore, the curvature perturbation is not generally conserved on superhorizon scales. This feature distinguishes two-field inflation from the simplest single-field models.

Two-field inflation also produces richer observational signatures. Curvature–entropy conversion can modify the primordial power spectrum and generate correlations between adiabatic and isocurvature perturbations \cite{Wands:2002bn,GrootNibbelink:2001qt,Wands:2007bd,Vernizzi:2006ve,Lyth:2005fi}. Nonlinear conversion can generate primordial non-Gaussianity \cite{Byrnes:2008wi,Byrnes:2010em,Peterson:2010mv,Elliston:2011et,Anderson:2012em,Elliston:2012wm,Gao:2008dt,Arroja:2008yy,Langlois:2008mn}. General kinetic terms and curved field-space geometries further modify these effects \cite{Langlois:2008wt,Langlois:2008qf,Gong:2011uw,Peterson:2011yt,Chen:2009zp}. Heavy fields may also leave observable signatures when the inflationary trajectory turns in field space \cite{Tolley:2009fg,Achucarro:2010jv,Achucarro:2010da,Achucarro:2012sm,Achucarro:2012yr,Gao:2012uq}. Such effects can change the effective sound speed, generate features in the power spectrum, and modify non-Gaussian correlations. Field-space curvature can even affect the stability of the inflationary trajectory \cite{Renaux-Petel:2015mga}. These developments show that two-field inflation is not simply a single-field model with one additional scalar. The coupling between the two degrees of freedom changes the dynamics of the primordial perturbations.

The quantum-state structure is also richer. A single scalar field naturally leads to a two-mode squeezed state. In contrast, two coupled scalar fields contain four quantum modes, \((1,{k})\), \((1,-{k})\), \((2,{k})\), and \((2,-{k})\). It has been shown that such a system is naturally described by four-mode squeezed states associated with the symplectic group \(Sp(4,\mathbb{R})\) \cite{Colas:2021llj}. The state contains squeezing as well as quantum transfer between the two field sectors. This provides the natural quantum-state framework for studying complexity in two-field inflation.

This structure gives a new physical motivation for Krylov complexity. In single-field inflation, Krylov spreading is mainly generated by the time-dependent background and the squeezing of the \(({k},-{k})\) pair. In two-field inflation, curvature–entropy coupling introduces an additional source of quantum evolution. Mode mixing, trajectory turning, entropy masses, and field-space geometry can modify the Hamiltonian and open additional channels for squeezing and operator growth. Krylov complexity may therefore separate background-induced quantum spreading from interaction-induced spreading. It may also provide information that is complementary to the power spectrum, isocurvature fraction, and non-Gaussianity. In this sense, Krylov complexity provides a possible quantum-information probe of genuine multi-field dynamics.

In the diagonal squeezing basis used below, this four-mode structure leads to a two-index occupation basis \(|n_k,n_{-k},m_k,m_{-k}\rangle\). It therefore provides two directions for Krylov-space spreading rather than the single ladder of the usual two-mode squeezed state. The same four-mode framework can also be used to study reduced-state decoherence when one sector is traced out \cite{Colas:2021llj}, although the present analysis focuses on unitary Krylov-space dynamics.

Observational constraints make this question especially interesting. Current CMB measurements strongly constrain primordial isocurvature perturbations, but they do not imply that additional fields were absent during inflation. Extra fields can be heavy, can decay, or can transfer their perturbations into the curvature sector before the observable epoch. Current CMB observations place strong constraints on surviving isocurvature contributions while still allowing nontrivial multi-field dynamics during inflation \cite{Planck:2018jri}. Krylov complexity probes the dynamical evolution of the quantum state rather than only its final correlation functions. It may therefore provide a different theoretical diagnostic of multi-field interactions. The two-field model also provides the simplest setting for extensions to general multi-field inflation, where several entropy directions and transfer channels may be present \cite{Dimopoulos:2005ac,Tye:2008ef,Seery:2012vj,Kaiser:2012ak,Kaiser:2013sna,Easther:2013rva,Dias:2015rca,Dias:2016slx,Dias:2016rjq,Dias:2017gva,Kaiser:2010ps,Cespedes:2012hu,Shiu:2011qw}. These studies show that systems with three, tens, or even hundreds of fields can exhibit dynamics that cannot always be reduced to a two-field description. 

In this work, we study Krylov-space dynamics in two-field inflation. We solve the coupled curvature and entropy perturbations, construct the quadratic Hamiltonian, and describe the resulting Gaussian state in terms of two squeezing channels. We then evaluate the squeezing parameters, Krylov complexity, Krylov entropy, and channel-resolved Lanczos coefficients. The main purpose is to determine how curvature–entropy coupling modifies Krylov-space spreading across horizon crossing. In contrast to the effectively single-field case, the two-field system contains two squeezing eigenchannels with different dynamical responses. Particular attention is paid to their different onset times and to the oscillatory behavior of their Krylov complexities before and after horizon exit. The comparison of the two channels allows the effects of multi-field coupling on squeezing, Krylov spreading, and hopping strengths to be identified directly.

This paper is organized as follows. In Sec. \ref{sec:multi‑field inflation}, we introduce the two-field inflationary model and construct the quadratic Hamiltonian for the coupled curvature and entropy perturbations. In Sec. \ref{sec:Evolution Equations of the squeezing Parameters}, we formulate the four-mode squeezed state and analyze the evolution of the two squeezing channels. In Sec. \ref{sec:Krylov complexity of two-field inflation}, we derive the Krylov complexity of the two-field system. In Sec. \ref{sec:Krylov entropy and Lanczos coefficients}, we study the Krylov entropy and the channel-resolved Lanczos coefficients. Finally, Sec. \ref{sec:Conclusions and Outlook} summarizes our main results and discusses possible extensions to general multi-field inflation.

\section{Quantum Hamiltonian in terms of creation-annihilation for two-field inflation}
\label{sec:multi‑field inflation}

We work within the framework of two‑field inflation, which is sufficiently generic to shed light on the evolution of quantum complexity in more general models. The canonical evolution of multi‑field systems in the adiabatic basis has been investigated in Ref. \cite{Nilles:2001fg}. We work in the Bunch–Davies basis and solve the coupled curvature and entropy perturbations numerically. This formulation allows the effects of curvature–entropy coupling to be tracked directly through the mode functions and the quadratic Hamiltonian.

In two‑field inflation, the quadratic action governing the dynamics of linear perturbations takes the following form \cite{GrootNibbelink:2001qt}
\begin{equation}\label{eq:s of two‑field inflation}
 S=\int d^{3}x d\tau\, a^{2}\left[\epsilon\left(\xi'-\sqrt{2/\epsilon}\,\Omega a\psi\right)^{2}
-\epsilon\left(\nabla\xi\right)^{2}
+\frac12 \psi'^{2}
-\frac12\left(\nabla\psi\right)^{2}
-\frac12 a^{2}\mu^{2}\psi^{2}\right]
\end{equation}
where \(\xi\) denotes the primordial comoving curvature perturbation, and \(\psi\) is the isocurvature perturbation describing fluctuation components orthogonal to the background inflationary trajectory in field‑space. In this expression, \(\mu\) is the effective mass of the entropy mode, and \(\Omega\) characterizes the turning rate of the background inflationary trajectory in field‑space.

For notational convenience, we introduce a dimensionless coupling parameter $\lambda$ defined as
\begin{equation}
\lambda(\tau)\equiv \frac{2\Omega(\tau)}{H}.
\end{equation}
The covariant time‑derivative $D_\tau\xi$ entering Eqs. \eqref{dynamical equation of xi} and \eqref{dynamical equation of psi} corresponds to a specific composite field combination
\begin{equation}
D_\tau\xi \equiv \xi' - aH\frac{\lambda}{\sqrt{2\epsilon}}\psi.
\end{equation}
Eqs. \eqref{dynamical equation of xi} and \eqref{dynamical equation of psi} govern the mutual linear‑order interactions between the curvature component $\xi$ and the entropy component $\psi$. Their mutual coupling is governed by $\lambda$, which is allowed to vary arbitrarily with cosmic time according to the dynamical evolution of the background trajectory within field‑space.

Using the covariant derivative defined above, the quadratic action can be rewritten as
\begin{equation}
    S = \int d^3x\,d\tau\, a^2\left[\epsilon(D_\tau\xi)^2 - \epsilon(\nabla\xi)^2 + \frac12\psi'^2 - \frac12(\nabla\psi)^2 - \frac12 a^2\mu^2\psi^2\right].
\end{equation}
We then vary the action with respect to \(\xi\) and \(\psi\), respectively. Using the relation \(\frac{(a^2\epsilon)'}{a^2\epsilon} = aH(2+\eta_{\rm SR})\) and the Fourier transformation \(-\nabla^2 \to k^2\). The dynamical equations obeyed by the perturbation modes follow from the action \eqref{eq:s of two‑field inflation}:
\begin{equation}\label{dynamical equation of xi}
(D_\tau\xi)' + aH(2+\eta_{\rm SR}) D_\tau\xi-\nabla^2\xi = 0,
\end{equation}
\begin{equation}\label{dynamical equation of psi}
\psi'' + 2aH\psi' -\nabla^2\psi + a^2\mu^2\psi + aH\sqrt{2\epsilon}\,\lambda D_\tau\xi = 0.
\end{equation}
The canonical momenta associated with \(\xi\) and \(\psi\) are
\begin{equation}
\pi_\xi = 2 a^2 \epsilon D_\tau \xi,\quad \pi_\psi = a^2 \psi'.
\end{equation}
We then perform the Legendre transform to obtain the Hamiltonian
\begin{equation}\label{Hamiltonian of two field}
    \mathcal{H}=\int_{x}\left[\frac{1}{4a^{2}\epsilon}\Pi_{\xi}^{2}
+a^{2}\epsilon(\nabla\xi)^{2}
+\frac{1}{2a^{2}}\Pi_{\psi}^{2}
+\frac{a^{2}}{2}(\nabla\psi)^{2}
+\frac{1}{2}a^{4}\mu^{2}\psi^{2}
+\frac{aH}{2}\frac{\lambda}{\sqrt{2\epsilon}}\big(\Pi_{\xi}\psi+\psi\Pi_{\xi}\big)\right].
\end{equation}
It is worth explicitly emphasizing the final term
\begin{equation}
 \mathcal  H_{\rm mix} \propto \lambda\big(\Pi_\xi \psi + \psi \Pi_\xi\big),
\end{equation}
which represents the most direct manifestation of the two-field coupling in the Hamiltonian. 

Since the system consists of mutually‑coupled differential equations, the field operators $\hat{\xi}(\tau,\boldsymbol{x})$, $\hat{\psi}(\tau,\boldsymbol{x})$ together with their corresponding conjugate momenta need to be decomposed via creation‑annihilation ladder operators within Fourier‑mode representation, according to
\begin{equation}\label{eq:ladder operators}
    \begin{split}
   &  \hat{\xi}(\tau,\boldsymbol x)
=\int\frac{d^3\boldsymbol k}{(2\pi)^3} e^{i\boldsymbol k\cdot\boldsymbol x}\Big[
\xi_1(\tau,k)\,\hat a_1(\boldsymbol k)
+\xi_2(\tau,k)\,\hat a_2(\boldsymbol k)
+\xi_1^*(\tau,k)\,\hat a_1^\dagger(-\boldsymbol k)
+\xi_2^*(\tau,k)\,\hat a_2^\dagger(-\boldsymbol k)
\Big],
\\&
\hat{\psi}(\tau,\boldsymbol x)
=\int\frac{d^3\boldsymbol k}{(2\pi)^3} e^{i\boldsymbol k\cdot\boldsymbol x}\Big[
\psi_1(\tau,k)\,\hat a_1(\boldsymbol k)
+\psi_2(\tau,k)\,\hat a_2(\boldsymbol k)
+\psi_1^*(\tau,k)\,\hat a_1^\dagger(-\boldsymbol k)
+\psi_2^*(\tau,k)\,\hat a_2^\dagger(-\boldsymbol k)
\Big],
\\&
\hat{\pi}_\xi(\tau,\boldsymbol x)
=\int\frac{d^3\boldsymbol k}{(2\pi)^3} e^{i\boldsymbol k\cdot\boldsymbol x}\Big[
\pi_1^\xi(\tau,k)\,\hat a_1(\boldsymbol k)
+\pi_2^\xi(\tau,k)\,\hat a_2(\boldsymbol k)
+{\pi_1^\xi}^*(\tau,k)\,\hat a_1^\dagger(-\boldsymbol k)
+{\pi_2^\xi}^*(\tau,k)\,\hat a_2^\dagger(-\boldsymbol k)
\Big],
\\&
\hat{\pi}_\psi(\tau,\boldsymbol x)
=\int\frac{d^3\boldsymbol k}{(2\pi)^3} e^{i\boldsymbol k\cdot\boldsymbol x}\Big[
\pi_1^\psi(\tau,k)\,\hat a_1(\boldsymbol k)
+\pi_2^\psi(\tau,k)\,\hat a_2(\boldsymbol k)
+{\pi_1^\psi}^*(\tau,k)\,\hat a_1^\dagger(-\boldsymbol k)
+{\pi_2^\psi}^*(\tau,k)\,\hat a_2^\dagger(-\boldsymbol k)
\Big],
    \end{split}
\end{equation}
where \(\hat a^\dagger_{a}(\boldsymbol{k})\) and \(\hat a_{a}(\boldsymbol{k})\) (\(a=1,2\)) are the creation and annihilation operators, which satisfy the commutation relation
\begin{equation}
    \big[\hat{a}_{a}(\boldsymbol{k}),\hat{a}_{b}^{\dagger}(\boldsymbol{k}')\big]=(2\pi)^3\delta_{ab}\delta^{(3)}(\boldsymbol{k}-\boldsymbol{k}'),
\end{equation}
and ensure that the field and momentum operators obey the standard equal‑time commutation relations
\begin{equation}
\big[\hat{\xi}(\tau,\boldsymbol{x}),\hat{\pi}_\xi(\tau,\boldsymbol{y})\big]=i\delta^{(3)}(\boldsymbol{x}-\boldsymbol{y}),\quad   \big[\hat{\psi}(\tau,\boldsymbol{x}),\hat{\pi}_\psi(\tau,\boldsymbol{y})\big]=i\delta^{(3)}(\boldsymbol{x}-\boldsymbol{y}).
\end{equation}
Furthermore, the dimensionless power spectra are given by
\begin{equation}
    \begin{split}
       & \mathcal{P}_\xi(k,\tau) = \frac{k^3}{2\pi^2}\left(|\xi_{1k}|^2 + |\xi_{2k}|^2\right),
       \\&
       \mathcal{P}_\psi(k,\tau) = \frac{k^3}{2\pi^2}\left(|\psi_{1k}|^2 + |\psi_{2k}|^2\right).
    \end{split}
\end{equation}
To determine the quantum evolution of the coupled perturbations, we first solve the mode equations numerically. We impose the Bunch-Davies initial condition deep inside the horizon. Two linearly independent initial modes are considered. Their evolution determines the four mode functions \(\xi_{1k}\), \(\xi_{2k}\), \(\psi_{1k}\), and \(\psi_{2k}\). These mode functions provide the basic input for constructing the quadratic Hamiltonian and the squeezed state. 

For the numerical analysis, we set the Hubble parameter to \(H=1\) and choose the slow-roll parameters \(\epsilon=0.002\) and \(\eta_{\rm SR}=0.01\). The effective mass of the entropy mode and the curvature–entropy coupling are taken as \(\mu=0.5H\) and \(\lambda=0.1\), respectively. The reference comoving wavenumber is fixed at \(k=0.01\), and the evolution is solved over the conformal-time interval \(\tau_i=-10000\) to \(\tau_f=-0.02\). To characterize the evolution relative to the Hubble scale, we introduce the dimensionless variable $x=-\log(-k\tau)$. In a quasi-de Sitter background, \(aH\simeq-1/\tau\), so that \(-k\tau\simeq k/(aH)\). Therefore, \(x\) measures the logarithmic ratio between the Hubble scale and the comoving mode scale. The condition \(x=0\) corresponds to \(k=aH\), namely horizon crossing, while \(x<0\) and \(x>0\) characterize the subhorizon and superhorizon regimes, respectively. Since inflation continuously stretches a fixed comoving mode relative to the Hubble radius, \(x\) provides an e-fold-like variable that naturally tracks the transition from quantum oscillations on subhorizon scales to squeezing and superhorizon evolution. The logarithmic form is also useful because inflation involves a large hierarchy of physical scales, which can be represented within a compact dimensionless variable.

\begin{figure}[htbp]
    \centering
    \begin{minipage}{0.5\textwidth}
        \centering
        \includegraphics[width=\linewidth]{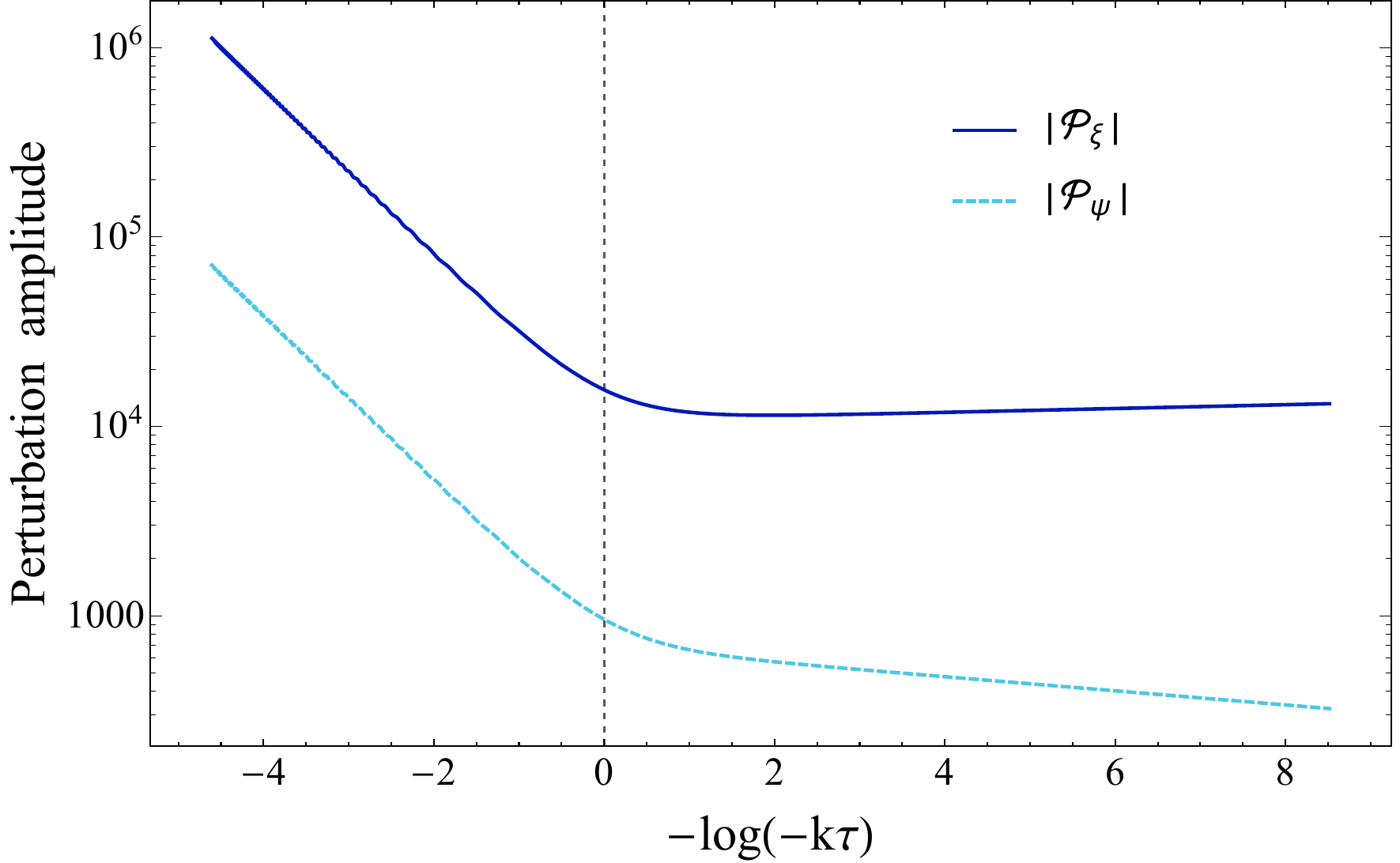}
    \end{minipage}

    \vspace{1em}
    
    \begin{minipage}{0.45\textwidth}
        \centering
        \includegraphics[width=\linewidth]{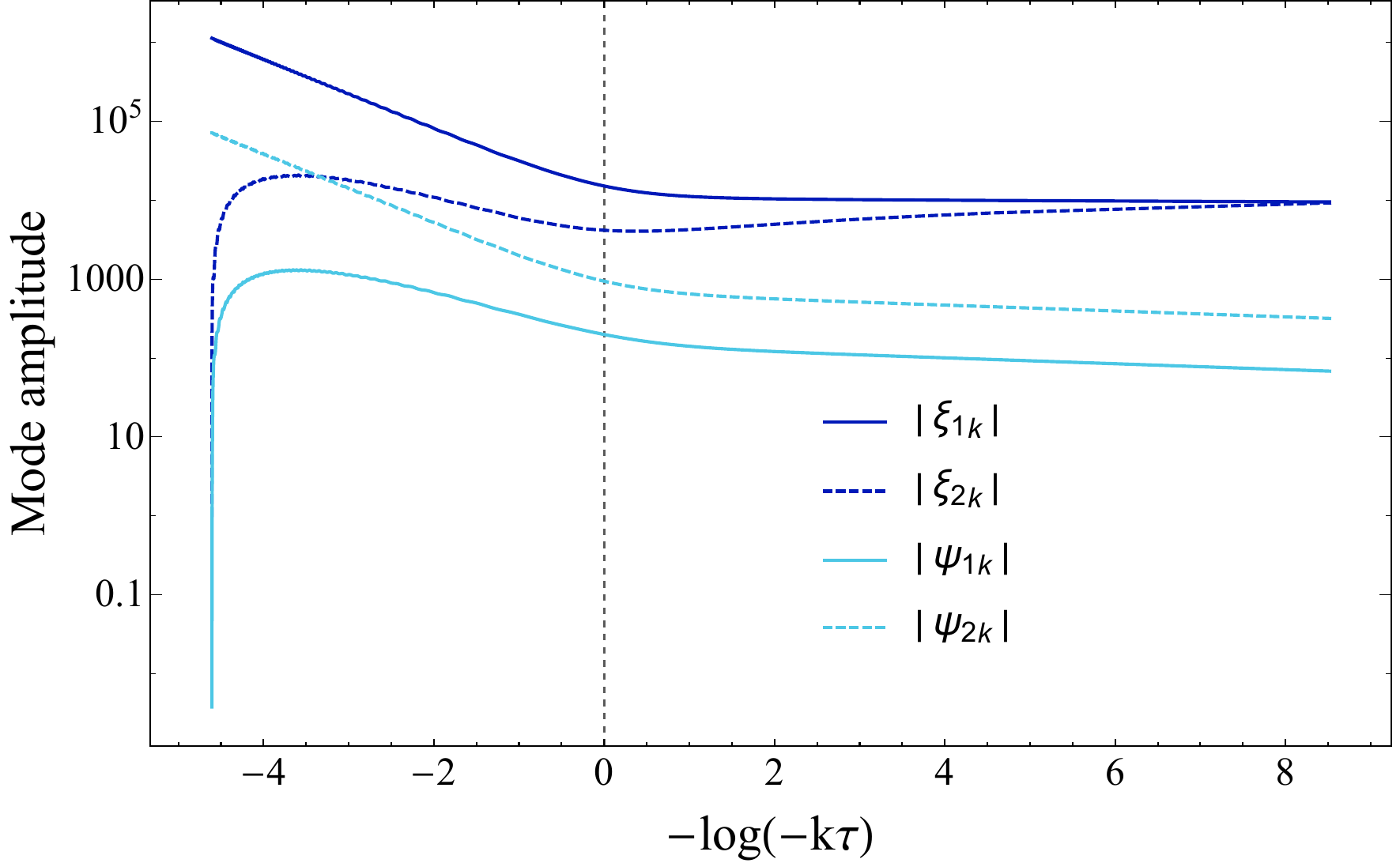}
    \end{minipage}
    \hfill
    \begin{minipage}{0.45\textwidth}
        \centering
         \includegraphics[width=\linewidth]{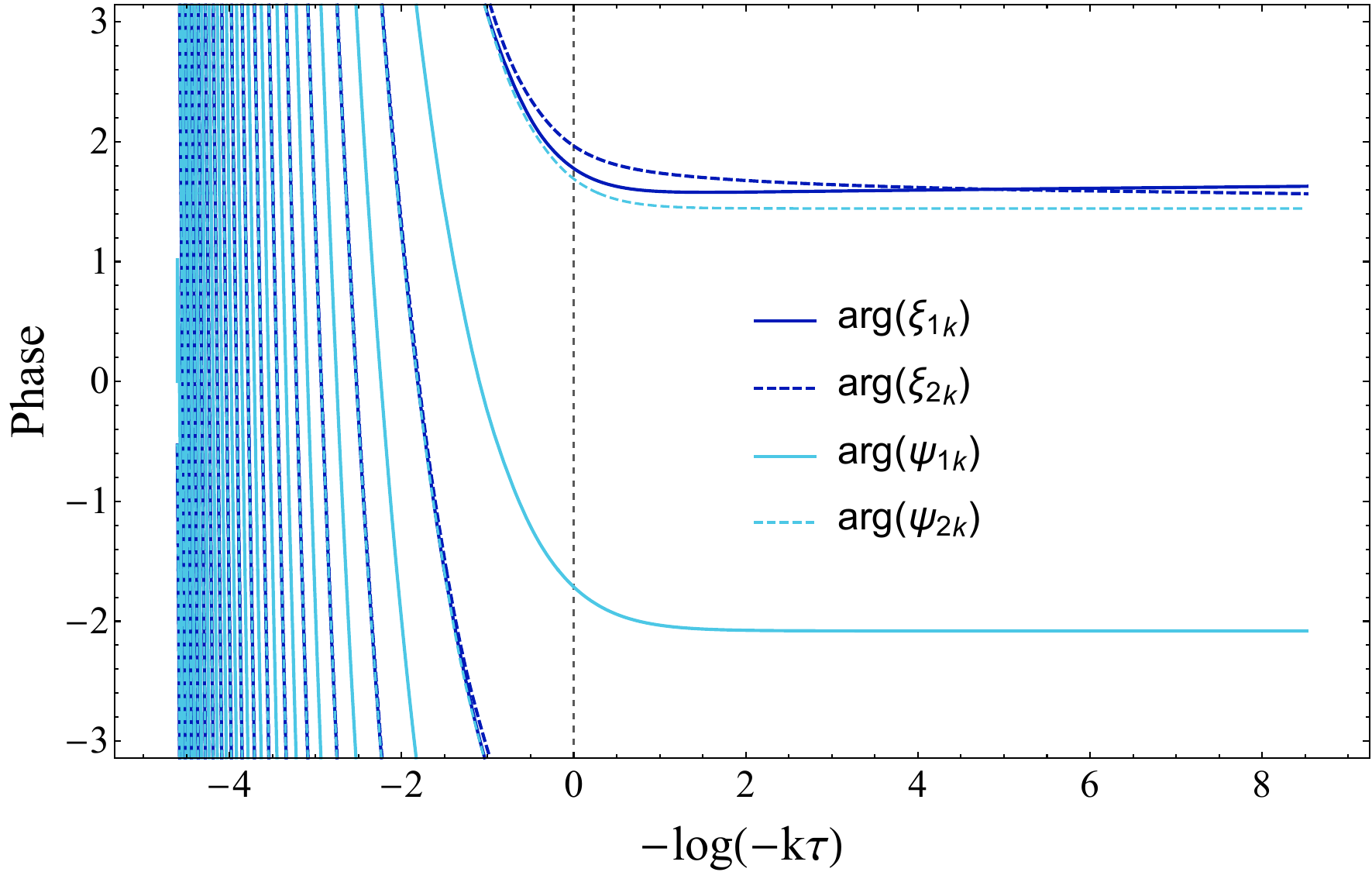}
    \end{minipage}
    \caption{Evolution of the perturbation amplitudes, mode-function amplitudes, and phases during two-field inflation as functions of \(x=-\log(-k\tau)\). The vertical dotted line at \(x=0\) denotes horizon crossing. The upper panel shows the total curvature and entropy perturbation amplitudes. The lower-left panel shows \(|\xi_{1k}|\), \(|\xi_{2k}|\), \(|\psi_{1k}|\), and \(|\psi_{2k}|\), while the lower-right panel shows their phases. Physical evolution proceeds from left to right, with \(x<0\) and \(x>0\) corresponding to the subhorizon and superhorizon regimes, respectively.}
    \label{fig:perturbation}
\end{figure}

Fig. \ref{fig:perturbation} shows the evolution of the coupled curvature and entropy perturbations. Deep inside the horizon, the mode functions exhibit rapid phase evolution. The curvature–entropy coupling generates nonzero mixed components, reflecting the transfer between the adiabatic and entropy sectors. After horizon crossing, the curvature perturbation approaches a slowly varying amplitude, while the entropy contribution is suppressed. The phases also become much less rapidly varying. The apparent jumps in the phase curves arise from the principal branch of the complex argument and do not represent physical discontinuities. This behavior is consistent with the standard multi-field picture in which entropy perturbations can source curvature perturbations when the background trajectory bends in field space \cite{Gordon:2000hv}. The suppression of rapid phase evolution after horizon exit is also closely related to the squeezing of inflationary perturbations  \cite{Albrecht:1992kf}.

Inserting \eqref{eq:ladder operators} into the Hamiltonian \eqref{Hamiltonian of two field}, one finds
\begin{equation}\label{eq:Hamiltonian of two mode with creation‑annihilation ladder operators}
\hat{\mathcal{H}_\mathrm{k}}=\frac12\sum_{bc}\int_{k}\Big[\mathcal{A}_{bc}\hat{a}_{b}(\boldsymbol{k})\hat{a}_{c}(-\boldsymbol{k})
+\mathcal{B}_{bc}\hat{a}_{b}^{\dagger}(\boldsymbol{k})\hat{a}_{c}(\boldsymbol{k})
+\mathcal{B}_{bc}^{*}\hat{a}_{b}(\boldsymbol{k})\hat{a}_{c}^{\dagger}(\boldsymbol{k})
+\mathcal{A}_{bc}^{*}\hat{a}_{b}^{\dagger}(\boldsymbol{k})\hat{a}_{c}^{\dagger}(-\boldsymbol{k})\Big],
\end{equation}
with $\mathcal{A}_{ab}$ and $\mathcal{B}_{ab}$ given by
\begin{align}
{A}_{ab}
&=\frac{\pi_{a}^{\xi}\pi_{b}^{\xi}}{2a^{2}\epsilon}
+\frac{\pi_{a}^{\psi}\pi_{b}^{\psi}}{a^{2}}
+2a^{2}\epsilon k^{2}\xi_{a}\xi_{b}
+a^{2}\bigl(k^{2}+a^{2}\mu^{2}\bigr)\psi_{a}\psi_{b}
+\frac{aH\lambda}{\sqrt{2\epsilon}}\bigl(\pi_{a}^{\xi}\psi_{b}+\psi_{a}\pi_{b}^{\xi}\bigr),
\\[4pt]
{B}_{ab}
&=\frac{\pi_{a}^{\xi*}\pi_{b}^{\xi}}{2a^{2}\epsilon}
+\frac{\pi_{a}^{\psi*}\pi_{b}^{\psi}}{a^{2}}
+2a^{2}\epsilon k^{2}\xi_{a}^{*}\xi_{b}
+a^{2}\bigl(k^{2}+a^{2}\mu^{2}\bigr)\psi_{a}^{*}\psi_{b}
+\frac{aH\lambda}{\sqrt{2\epsilon}}\bigl(\pi_{a}^{\xi*}\psi_{b}+\psi_{a}^{*}\pi_{b}^{\xi}\bigr),
\end{align}
where, $A_{ab}$ controls pair creation/squeezing, while $B_{ab}$ contains number-conserving evolution and mode mixing. In the original mode basis, the pair-production matrix \(A_{ij}\) is generally non-diagonal. In the present analysis, we focus on the two dominant diagonal squeezing channels. The off-diagonal coefficient \(A_{12}\) characterizes pair production involving both channels and the present analysis focuses on the two diagonal squeezing channels, while the effects of off-diagonal pair production are not included explicitly. The two-channel description is adopted when the off-diagonal contribution remains subdominant. The curvature–entropy coupling is nevertheless retained in the coupled mode functions and in the diagonal Hamiltonian coefficients. We therefore explicitly expand the diagonal coefficients associated with the two squeezing channels, corresponding to \(b=c=1\) and \(b=c=2\). 

\begin{figure}[htbp]
    \centering
    \begin{minipage}{0.5\textwidth}
        \centering
        \includegraphics[width=\linewidth]{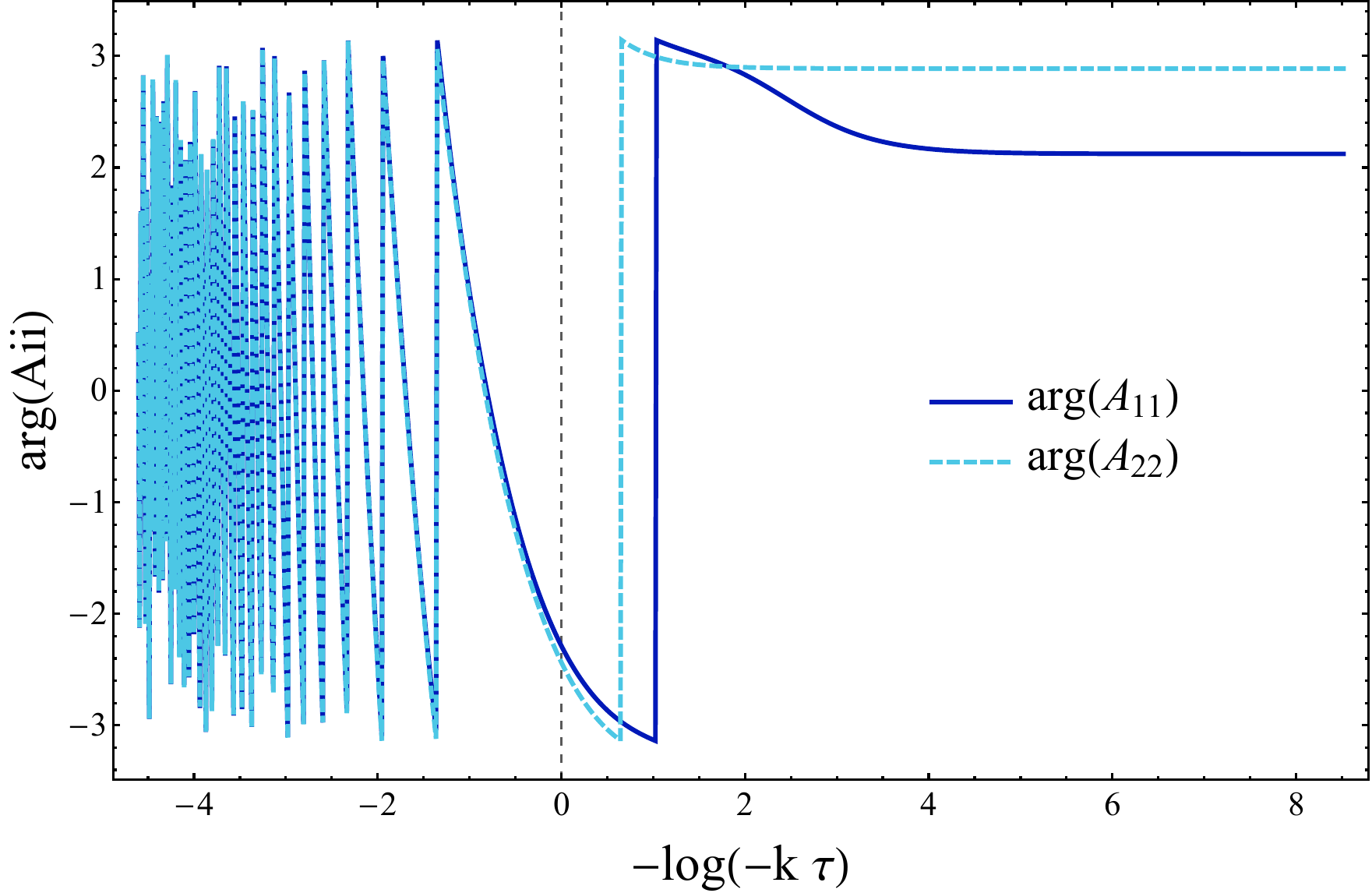}
    \end{minipage}

    \vspace{1em}
    
    \begin{minipage}{0.45\textwidth}
        \centering
        \includegraphics[width=\linewidth]{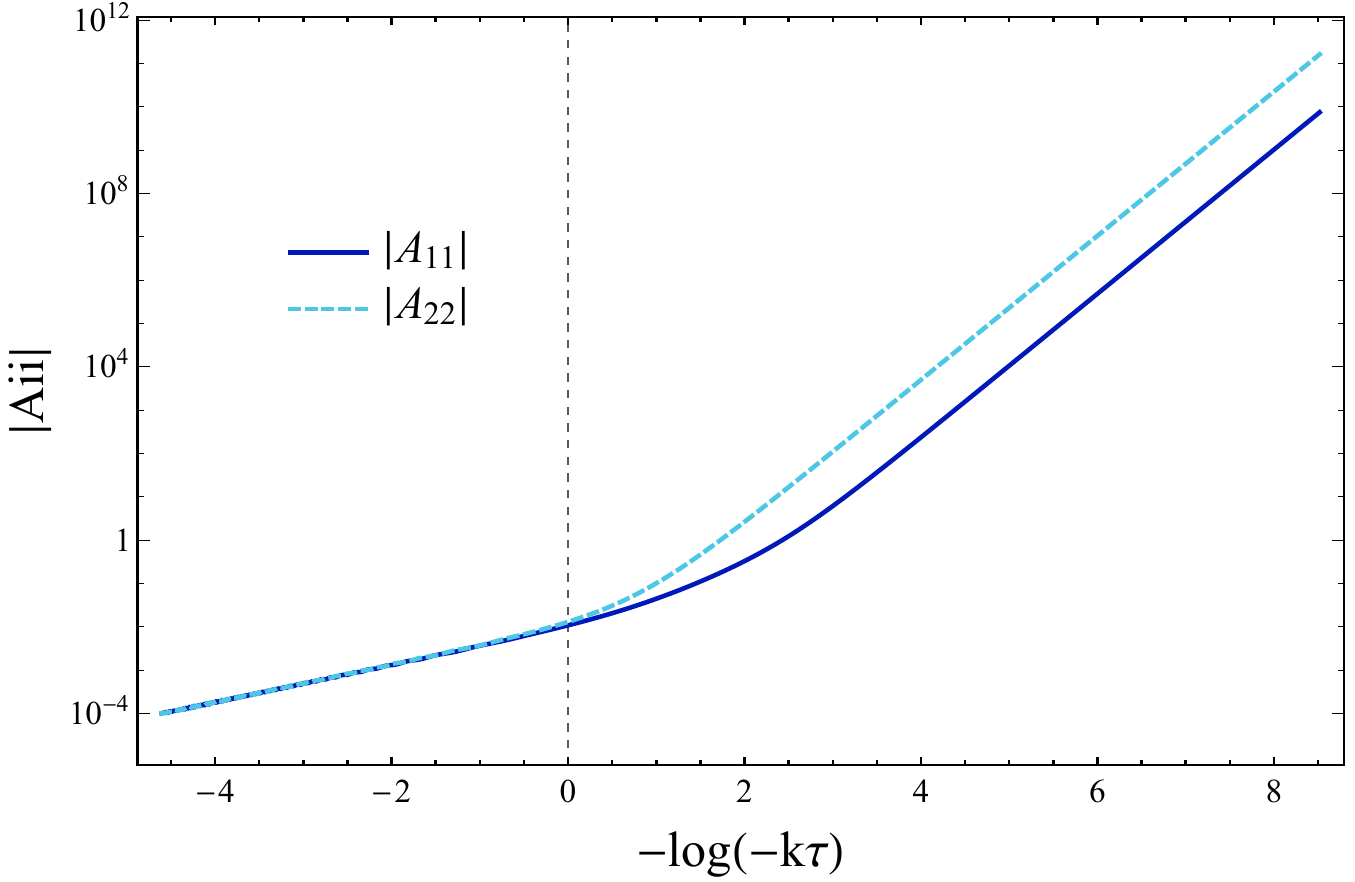}
    \end{minipage}
    \hfill
    \begin{minipage}{0.45\textwidth}
        \centering
         \includegraphics[width=\linewidth]{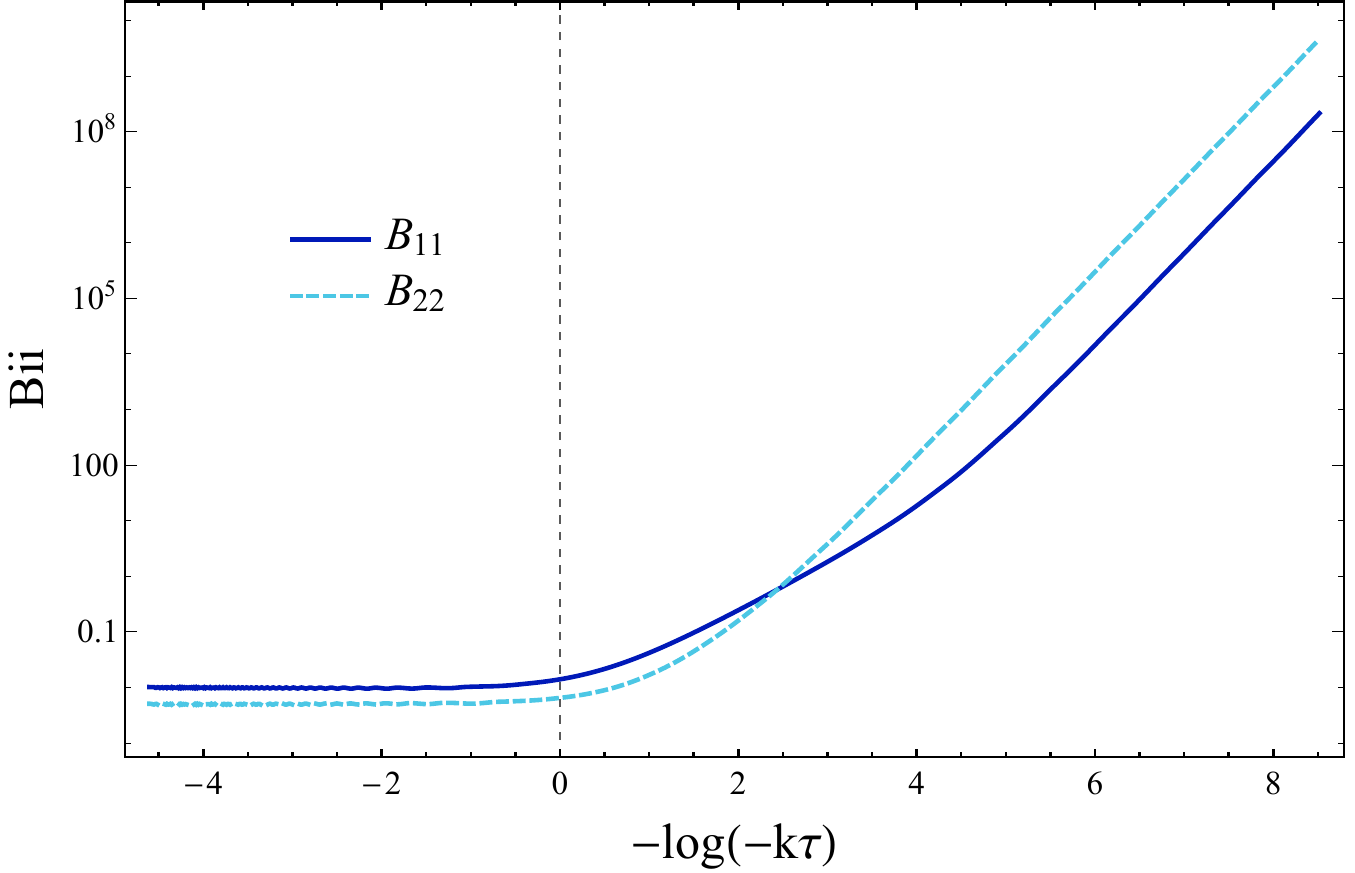}
    \end{minipage}
    \caption{Evolution of the Hamiltonian coefficients \(A_{11}\), \(A_{22}\), \(B_{11}\), and \(B_{22}\) as functions of \(x=-\log(-k\tau)\). The upper panel shows \(\arg A_{11}\) and \(\arg A_{22}\). The lower-left panel shows \(|A_{11}|\) and \(|A_{22}|\), while the lower-right panel shows \(B_{11}\) and \(B_{22}\). The vertical dotted line at \(x=0\) denotes horizon crossing.}
    \label{fig:ABII}
\end{figure}

The pair-production coefficient for the first channel reads
\begin{equation}
    \begin{split}
    A_{11}  & = 2a^2\varepsilon\left(\xi_{1k}'-aH\lambda \psi_{1k}/\sqrt{2\varepsilon}\right)^2 + a^2(\psi_{1k}')^2 + 2a^2\varepsilon k^2 \xi_{1k}^2\\&+ a^2(k^2+a^2\mu^2)\psi_{1k}^2 + 2a^3H\lambda\sqrt{2\varepsilon}\left(\xi_{1k}' - aH\lambda \psi_{1k}/\sqrt{2\varepsilon}\right)\psi_{1k}
    \end{split}
\end{equation}
The number-conserving coefficient for the first channel reads
\begin{equation}
    \begin{split}
        B_{11}&=2a^2\varepsilon\left|\xi_{1k}' - aH\lambda \psi_{1k}/\sqrt{2\varepsilon}\right|^2 + a^2\left|\psi_{1k}'\right|^2 + 2a^2\varepsilon k^2\left|\xi_{1k}\right|^2\\&+ a^2(k^2+a^2\mu^2)\left|\psi_{1k}\right|^2
+2a^3H\lambda\sqrt{2\varepsilon} \mathrm{Re}\left[\left(\xi_{1k}' - aH\lambda \psi_{1k}/\sqrt{2\varepsilon}\right)^* \psi_{1k}\right].
    \end{split}
\end{equation}
The pair-production coefficient for the second channel reads
\begin{equation}
    \begin{split}
       A_{22}  & =2a^2\varepsilon\left(\xi_{2k}'-aH\lambda \psi_{2k}/\sqrt{2\varepsilon}\right)^2 + a^2(\psi_{2k}')^2 + 2a^2\varepsilon k^2 \xi_{2k}^2\\&+a^2(k^2+a^2\mu^2)\psi_{2k}^2 + 2a^3H\lambda\sqrt{2\varepsilon}\left(\xi_{2k}' - aH\lambda \psi_{2k}/\sqrt{2\varepsilon}\right)\psi_{2k}.
    \end{split}
\end{equation}
The number-conserving coefficient for the second channel reads
\begin{equation}
        \begin{split}
          B_{22}&=2a^2\varepsilon\left|\xi_{2k}' - aH\lambda \psi_{2k}/\sqrt{2\varepsilon}\right|^2 + a^2\left|\psi_{2k}'\right|^2 + 2a^2\varepsilon k^2\left|\xi_{2k}\right|^2\\&+ a^2(k^2+a^2\mu^2)\left|\psi_{2k}\right|^2
+ 2a^3H\lambda\sqrt{2\varepsilon} \mathrm{Re}\left[\left(\xi_{2k}' - aH\lambda \psi_{2k}/\sqrt{2\varepsilon}\right)^* \psi_{2k}\right].
    \end{split}
 \end{equation}
It can be observed that the two-field coupling does not vanish even when adopting two squeezing eigenchannels. It still explicitly enters the coefficients of each channel, embodied in the terms \(\xi_{1k}'-aH\lambda\psi_{1k}/\sqrt{2\varepsilon}\) and \(\xi_{2k}'-aH\lambda\psi_{2k}/\sqrt{2\varepsilon}\). Through \(\lambda\), the coupling modifies both the pair-production sector and the number-conserving sector simultaneously. Therefore, the two-field coupling enters the Krylov dynamics indirectly but systematically through the Hamiltonian coefficients. The coupling first modifies the coupled curvature–entropy mode functions and then changes \(A_{11},A_{22},B_{11}\), and \(B_{22}\). These coefficients determine the two squeezing histories, which in turn determine \(K_{1k}\), \(K_{2k}\), the Krylov entropy, and the channel-resolved Lanczos coefficients. Thus, the difference between the two Krylov channels is a dynamical consequence of curvature–entropy coupling rather than an independent assumption. The numerical evolution of these Hamiltonian coefficients is shown in Fig. \ref{fig:ABII}

Fig. \ref{fig:ABII} shows the Hamiltonian coefficients that determine the two squeezing channels. In the subhorizon regime, the phases of \(A_{11}\) and \(A_{22}\) vary rapidly, while their magnitudes remain relatively small. After horizon crossing, \(|A_{11}|\), \(|A_{22}|\), \(B_{11}\), and \(B_{22}\) increase significantly and the phase evolution becomes smoother. Since \(A_{ii}\) controls pair creation whereas \(B_{ii}\) contributes to number-conserving evolution, the different behavior of the two channels leads naturally to different squeezing histories.

\section{ squeezing Parameters of two-field inflation}
\label{sec:Evolution Equations of the squeezing Parameters}

To obtain the Fock‑space wave function corresponding to the target quantum state, we describe its dynamical evolution via the unitary operator \(\mathcal{U}_k = \hat{\mathcal{S}}_{\vec{k}}(r_k,\varphi_k)\hat{\mathcal{R}}_{\vec{k}}(\theta_k)\). Here, \(\hat{\mathcal{R}}_{\vec{k}}\) stands for the two‑mode rotation operator governed by the time‑dependent rotation angle \(\theta_k(\eta)\), whose explicit form reads
\begin{equation}\label{rotation operator}
\hat{\mathcal{R}}_{\vec{k}}(\theta_k) = \exp\left[-i\theta_k(\eta)\left(\hat{c}_{\vec{k}}\hat{c}_{\vec{k}}^\dagger + \hat{c}_{-\vec{k}}^\dagger\hat{c}_{-\vec{k}}\right)\right],
\end{equation}
while \(\hat{\mathcal{S}}_{\vec{k}}\) represents the two‑mode squeeze operator controlled by the squeeze parameter \(r_k(\eta)\) together with the squeeze angle \(\varphi_k(\eta)\). Its concrete mathematical expression is written as
\begin{equation}\label{squeezing operator}
\hat{\mathcal{S}}_{\vec{k}}(r_k,\varphi_k) = \exp\left[r_k(\eta)\left(e^{-2i\varphi_k(\eta)}\hat{c}_{\vec{k}}\hat{c}_{-\vec{k}} - e^{2i\varphi_k(\eta)}\hat{c}_{-\vec{k}}^\dagger\hat{c}_{\vec{k}}^\dagger\right)\right].
\end{equation}
When the evolution operator is ordered as \(\mathcal{U}_k = \hat{\mathcal{S}}_{\vec{k}}(r_k,\varphi_k)\hat{\mathcal{R}}_{\vec{k}}(\theta_k)\) and acts on the vacuum, \(\hat{\mathcal{R}}_{\vec{k}}(\theta_k)\) contributes only an overall phase to the initial vacuum state. Its dynamical phase rotation is subsequently encoded in the number-conserving coefficient \(\Omega_k\). Applying the squeeze operator \(\hat{\mathcal{S}}_{\vec{k}}(r_k,\varphi_k)\) \eqref{squeezing operator} onto the vacuum state yields the squeezing quantum state \cite{Albrecht:1992kf,Bolis:2018jmo}
\begin{equation}\label{eq:squeezed state}
|\psi\rangle_{sq}=\frac{1}{\cosh(r_k)}\sum_{n=0}^{\infty}(-1)^n e^{2in\varphi_k}\tanh^n r_k|n;n\rangle_{\vec{k},-\vec{k}}.
\end{equation}
For a fixed momentum pair \((\mathbf{k},-\mathbf{k})\), the two-field system contains four quantum modes, \((1,\mathbf{k})\), \((1,-\mathbf{k})\), \((2,\mathbf{k})\), and \((2,-\mathbf{k})\). The linear Gaussian evolution of such a system is described by the symplectic group \(Sp(4,\mathbb{R})\) \cite{Colas:2021llj}. A generic \(Sp(4,\mathbb{R})\) transformation contains both squeezing operations and passive rotations that mix the two field sectors. According to the Bloch–Messiah decomposition, the active squeezing block contains two independent squeezing parameters, while the remaining transformations describe rotations and inter-sector mixing. For the parameter range considered below, we work in the mode basis defined by the two independent coupled solutions and focus on the diagonal squeezing channels. The validity of this reduction is checked by the relative size of the off-diagonal Hamiltonian coefficients. The squeezing part of the evolution can therefore be written in terms of two two-mode squeezing operators,
\begin{equation}
    \hat S_{1k}(r_{1k},\phi_{1k}),\qquad
\hat S_{2k}(r_{2k},\phi_{2k}).
\end{equation}
This representation should be distinguished from the most general four-mode squeezed state written in the original field basis, where explicit transfer amplitudes between the two sectors also appear \cite{Colas:2021llj}. Importantly, diagonalizing the active squeezing block does not remove the physical curvature–entropy coupling. The coupling is encoded in the coupled mode functions and hence in $A_{11}$, $A_{22}$, $B_{11}$, and $B_{22}$. Therefore, $r_{1k}$ and $ r_{2k}$ describe two squeezing  eigen-channels of the same coupled system rather than two independent scalar fields.
\begin{equation}
    \begin{split}
        |\Psi_{ k}(\tau)\rangle = \hat{S}_{1 k}(r_{1 k},\varphi_{1 k}) \hat{S}_{2 k}(r_{2 k},\varphi_{2 k}) |0_{1, k},0_{1,- k},0_{2, k},0_{2,- k}\rangle,
    \end{split}
\end{equation}
\begin{equation}\label{eq:four-mode squeezed state}
    \begin{split}
   |\Psi_{ k}(\tau)\rangle  & = \frac{e^{i\chi_ k(\tau)}}{\cosh(r_{1 k})\cosh(r_{2 k})}\sum_{n=0}^{\infty}\sum_{m=0}^{\infty} (-1)^{n+m}
e^{2in\varphi_{1 k}} e^{2im\varphi_{2 k}}  \\
&\times \tanh^n(r_{1 k}) \tanh^m(r_{2 k})|n_{1, k},n_{1,- k},m_{2, k},m_{2,- k}\rangle.
    \end{split}
\end{equation}
The factorized form of Eq. \eqref{eq:four-mode squeezed state} reflects the diagonal active squeezing block; the underlying curvature–entropy coupling remains encoded in the time-dependent normal-mode functions and Hamiltonian coefficients.

The quantum state obeys the Schrödinger equation
\begin{equation}
    i\frac{\partial|\Psi_{k}(\tau)\rangle}{\partial\tau} = \hat{\mathcal{H}}_{k} |\Psi_{k}(\tau)\rangle.
\end{equation}
The operator actions for the first channel read
\begin{equation}
    \begin{split}
        &(\hat{N}_{1,{k}} + \hat{N}_{1,-{k}} + 1)|n,n,m,m\rangle = (2n+1)|n,n,m,m\rangle,
        \\&
        \hat{a}_1({k})\hat{a}_1(-{k})|n,n,m,m\rangle = n\,|n-1,n-1,m,m\rangle,
        \\&
\hat{a}_1^\dagger({k})\hat{a}_1^\dagger(-{k})|n,n,m,m\rangle = (n+1)\,|n+1,n+1,m,m\rangle.        
    \end{split}
\end{equation}
The second channel satisfies the corresponding relations
\begin{equation}
    \begin{split}
       & (\hat{N}_{2,{k}} + \hat{N}_{2,-{k}} + 1)|n,n,m,m\rangle = (2m+1)|n,n,m,m\rangle,
       \\&
       \hat{a}_2({k})\hat{a}_2(-{k})|n,n,m,m\rangle = m\,|n,n,m-1,m-1\rangle,
       \\&
       \hat{a}_2^\dagger({k})\hat{a}_2^\dagger(-{k})|n,n,m,m\rangle = (m+1)\,|n,n,m+1,m+1\rangle.
    \end{split}
\end{equation}
By comparing the coefficients of all \(|n,n,m,m\rangle\) components, the first and second squeezing eigenchannels each yield a set of evolution equations for $r_{k(1,2)}$ and \(\varphi_{k(1,2)}\).

\begin{equation}
    \begin{split}
  \frac{\mathrm{d}r_{1k}}{\mathrm{d}\tau}&  = \frac{1}{2}\Bigg|2a^2\varepsilon\big(\xi_{1k}'-a\mathrm{H}\lambda\psi_{1k}/\sqrt{(2\varepsilon)}\big)^2 + a^2(\psi_{1k}')^2 + 2a^2\varepsilon k^2\xi_{1k}^2
  \\&
+ a^2(k^2+a^2\mu^2)\psi_{1k}^2   + 2a^3\mathrm{H}\lambda\sqrt{(2\varepsilon)}\big(\xi_{1k}'-a\mathrm{H}\lambda\psi_{1k}/\sqrt{(2\varepsilon)}\big)\psi_{1k} \Bigg|
\\&
\times\sin\Bigg\{ 2\varphi_{1k}-\arg\bigg[ 2a^2\varepsilon\big(\xi_{1k}'-a\mathrm{H}\lambda\psi_{1k}/\sqrt{(2\varepsilon)}\big)^2+ 2a^2\varepsilon k^2\xi_{1k}^2
  \\& 
   + a^2(\psi_{1k}')^2  + a^2(k^2+a^2\mu^2)\psi_{1k}^2+ 2a^3\mathrm{H}\lambda\sqrt{(2\varepsilon)}\big(\xi_{1k}'-a\mathrm{H}\lambda\psi_{1k}/\sqrt{(2\varepsilon)}\big)\psi_{1k} \bigg] \Bigg\}.
    \end{split}
\end{equation}
\begin{equation}
    \begin{split}
        \frac{\mathrm{d}\varphi_{1k}}{\mathrm{d}\tau} & = 2a^2\varepsilon\big|\xi_{1k}'-a\mathrm{H}\lambda\psi_{1k}/\sqrt{(2\varepsilon)}\big|^2 + a^2|\psi_{1k}'|^2 + 2a^2\varepsilon k^2|\xi_{1k}|^2+ a^2(k^2+a^2\mu^2)|\psi_{1k}|^2 \\
& + 2a^3\mathrm{H}\lambda\sqrt{(2\varepsilon)}\mathrm{Re}\big[\big(\xi_{1k}'-a\mathrm{H}\lambda\psi_{1k}/\sqrt{(2\varepsilon)}\big)^*\psi_{1k}\big] \\
& + \frac12 \Bigg| 2a^2\varepsilon\big(\xi_{1k}'-a\mathrm{H}\lambda\psi_{1k}/\sqrt{(2\varepsilon)}\big)^2 + a^2(\psi_{1k}')^2 + 2a^2\varepsilon k^2\xi_{1k}^2+ a^2(k^2+a^2\mu^2)\psi_{1k}^2   \\
& + 2a^3\mathrm{H}\lambda\sqrt{(2\varepsilon)}\big(\xi_{1k}'-a\mathrm{H}\lambda\psi_{1k}/\sqrt{(2\varepsilon)}\big)\psi_{1k} \Bigg|\coth(2r_{1k}) \\
& \times \cos\Bigg\{ 2\varphi_{1k}-\arg\bigg[ 2a^2\varepsilon\big(\xi_{1k}'-a\mathrm{H}\lambda\psi_{1k}/\sqrt{(2\varepsilon)}\big)^2 + a^2(\psi_{1k}')^2 + 2a^2\varepsilon k^2\xi_{1k}^2 \\
& + a^2(k^2+a^2\mu^2)\psi_{1k}^2 + 2a^3\mathrm{H}\lambda\sqrt{(2\varepsilon)}\big(\xi_{1k}'-a\mathrm{H}\lambda\psi_{1k}/\sqrt{(2\varepsilon)}\big)\psi_{1k} \bigg] \Bigg\}.
    \end{split}
\end{equation}
\begin{equation}
    \begin{split}
        \frac{\mathrm{d}r_{2k}}{\mathrm{d}\tau} &=  \frac{1}{2}\Bigg|  2a^2\varepsilon\big(\xi_{2k}'-a\mathrm{H}\lambda\psi_{2k}/\sqrt{(2\varepsilon)}\big)^2 + a^2(\psi_{2k}')^2 + 2a^2\varepsilon k^2\xi_{2k}^2 \\
& + a^2(k^2+a^2\mu^2)\psi_{2k}^2+ 2a^3\mathrm{H}\lambda\sqrt{(2\varepsilon)}\big(\xi_{2k}'-a\mathrm{H}\lambda\psi_{2k}/\sqrt{(2\varepsilon)}\big)\psi_{2k} \Bigg| \\
& \times \sin\Bigg\{ 2\varphi_{2k}-\arg\bigg[ 2a^2\varepsilon\big(\xi_{2k}'-a\mathrm{H}\lambda\psi_{2k}/\sqrt{(2\varepsilon)}\big)^2 + 2a^2\varepsilon k^2\xi_{2k}^2  \\
& + a^2(\psi_{2k}')^2+ a^2(k^2+a^2\mu^2)\psi_{2k}^2+ 2a^3\mathrm{H}\lambda\sqrt{(2\varepsilon)}\big(\xi_{2k}'-a\mathrm{H}\lambda\psi_{2k}/\sqrt{(2\varepsilon)}\big)\psi_{2k} \bigg] \Bigg\}.
    \end{split}
\end{equation}
\begin{equation}
    \begin{split}
        \frac{\mathrm{d}\varphi_{2k}}{\mathrm{d}\tau}  &= 2a^2\varepsilon\big|\xi_{2k}'-a\mathrm{H}\lambda\psi_{2k}/\sqrt{(2\varepsilon)}\big|^2 + a^2|\psi_{2k}'|^2 + 2a^2\varepsilon k^2|\xi_{2k}|^2+ a^2(k^2+a^2\mu^2)|\psi_{2k}|^2 
    \\&  + 2a^3\mathrm{H}\lambda\sqrt{(2\varepsilon)}\mathrm{Re}\big[\big(\xi_{2k}'-a\mathrm{H}\lambda\psi_{2k}/\sqrt{(2\varepsilon)}\big)^*\psi_{2k}\big] \\
& + \frac12 \Bigg| 2a^2\varepsilon\big(\xi_{2k}'-a\mathrm{H}\lambda\psi_{2k}/\sqrt{(2\varepsilon)}\big)^2 + a^2(\psi_{2k}')^2 + 2a^2\varepsilon k^2\xi_{2k}^2+ a^2(k^2+a^2\mu^2)\psi_{2k}^2 \\
&  + 2a^3\mathrm{H}\lambda\sqrt{(2\varepsilon)}\big(\xi_{2k}'-a\mathrm{H}\lambda\psi_{2k}/\sqrt{(2\varepsilon)}\big)\psi_{2k} \Bigg|\coth(2r_{2k}) \\
& \times \cos\Bigg\{ 2\varphi_{2k}-\arg\bigg[ 2a^2\varepsilon\big(\xi_{2k}'-a\mathrm{H}\lambda\psi_{2k}/\sqrt{(2\varepsilon)}\big)^2 + a^2(\psi_2k')^2 + 2a^2\varepsilon k^2\xi_{2k}^2  \\
& + a^2(k^2+a^2\mu^2)\psi_{2k}^2+ 2a^3\mathrm{H}\lambda\sqrt{(2\varepsilon)}\big(\xi_{2k}'-a\mathrm{H}\lambda\psi_{2k}/\sqrt{(2\varepsilon)}\big)\psi_{2k} \bigg] \Bigg\}.
    \end{split}
\end{equation}
We find that the above four equations can be simplified into the following form 
\begin{equation}
    \begin{split}
       & ~~~~~~~~~~~~~\frac{dr_{1k}}{{d\tau}}=\frac{1}{2}|A_{11}|\sin\left(2\varphi_{1k}-\arg A_{11}\right),\\&
       \frac{d\varphi_{1k}}{{d\tau}}= B_{11}+\frac{1}{2}|A_{11}|\coth(2r_{1k})\cos\left(2\varphi_{1k}-\arg A_{11}\right),\\&
       ~~~~~~~~~~~~~~\frac{dr_{2k}}{{d\tau}} = \frac{1}{2} |A_{22}|\sin\left(2\varphi_{2k}-\arg A_{22}\right),\\&
 \frac{d\varphi_{2k}}{{d\tau}} = B_{22}+\frac{1}{2}|A_{22}|\coth(2r_{2k})\cos\left(2\varphi_{2k}-\arg A_{22}\right).
    \end{split}
\end{equation}
For the vacuum initial state, the squeezing strength satisfy \(r_{1k}=r_{2k}=0\). Direct integration of the equations for the squeezing angles is numerically inconvenient because of the factor \(\coth(2r_{ik})\). We therefore evolve the equivalent complex squeezing variables and reconstruct the squeezing strength and angles from the numerical solutions. This procedure allows the evolution to start from the vacuum initial condition without introducing an artificial nonzero initial squeezing strength.

\begin{figure}
    \centering
    \includegraphics[width=0.45\linewidth]{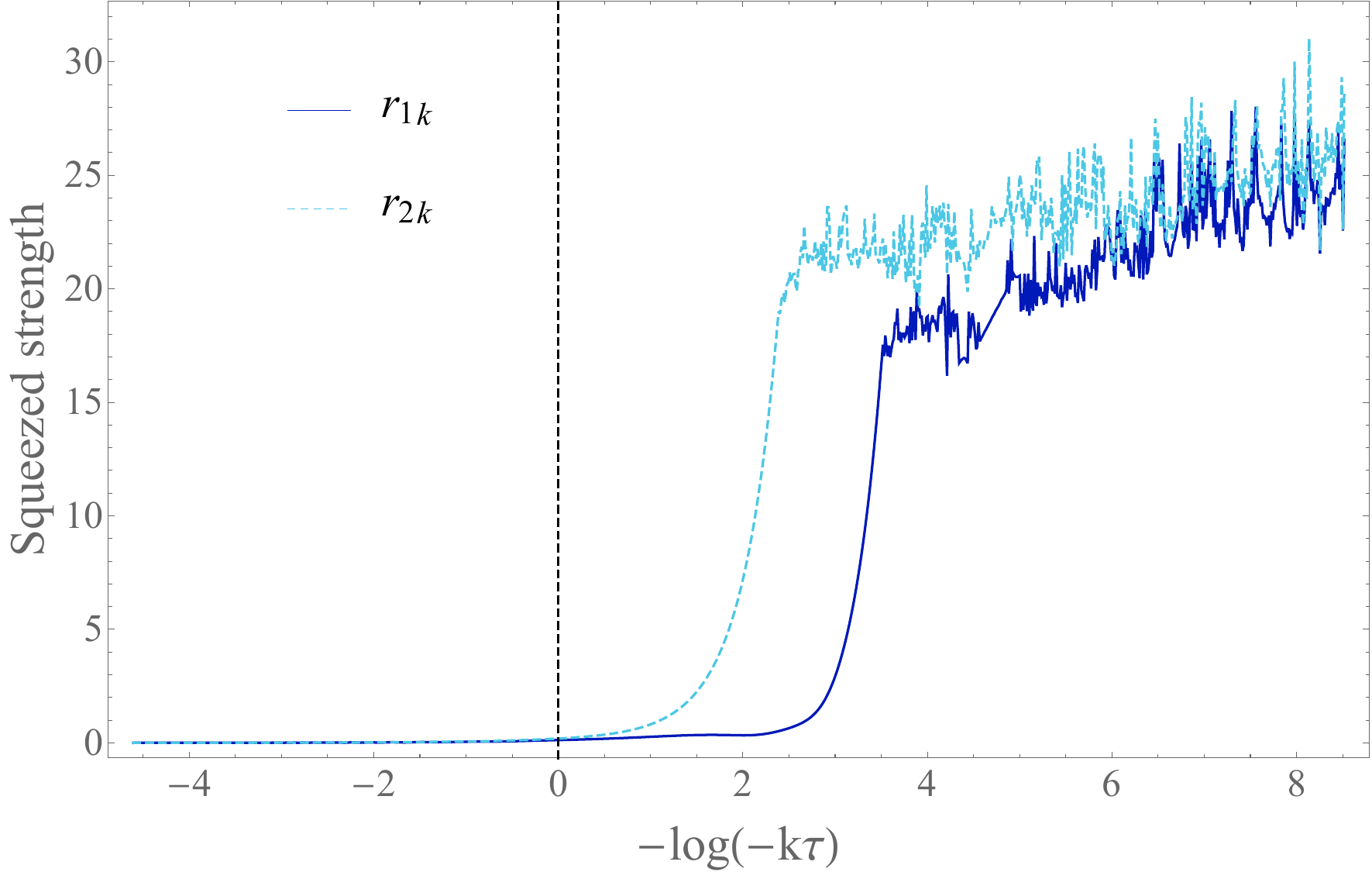}
        \includegraphics[width=0.45\linewidth]{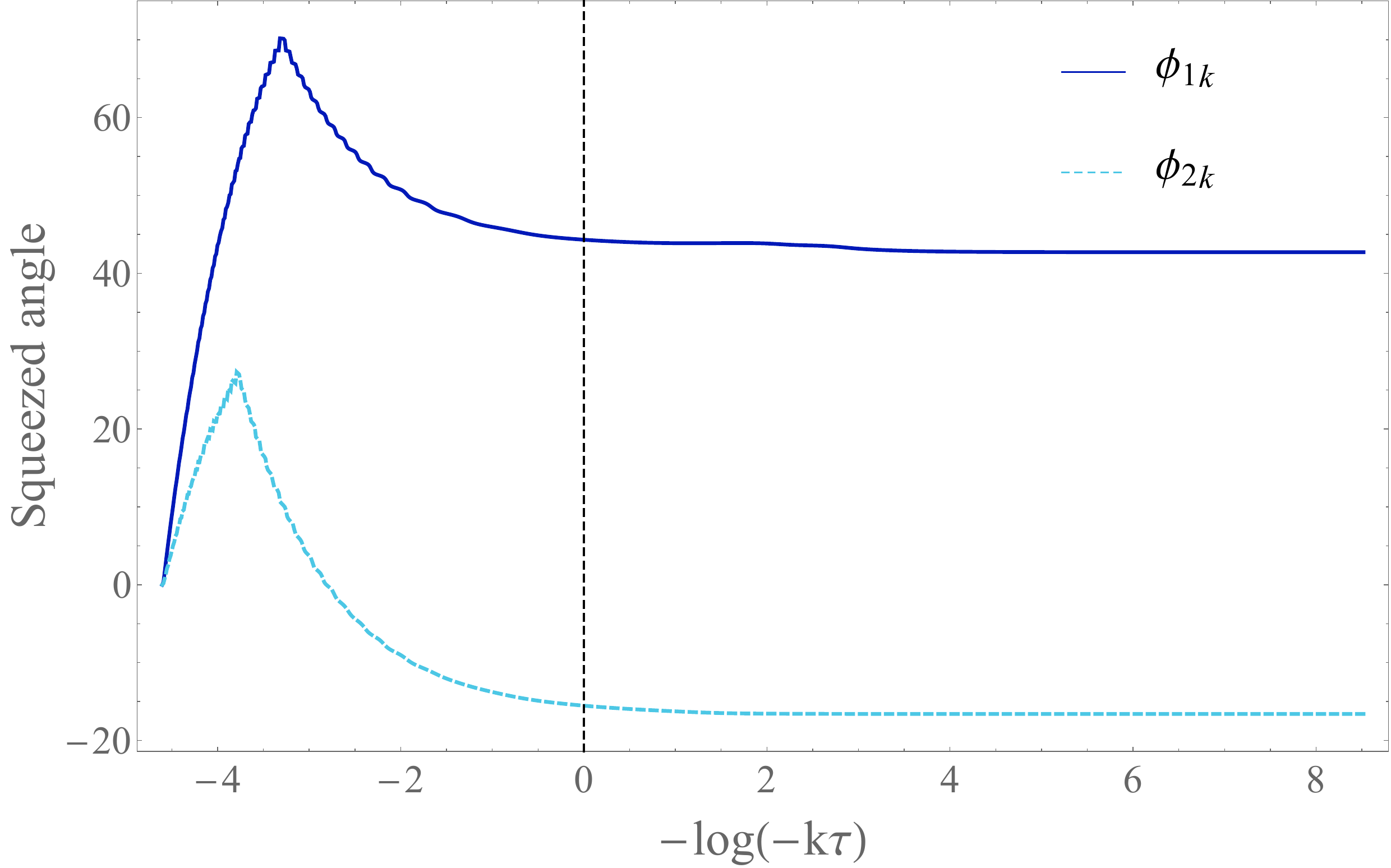}
    \caption{Evolution of the squeezing strength \(r_{1k}\) and \(r_{2k}\) (left panel) and the squeezing angles \(\varphi_{1k}\) and \(\varphi_{2k}\) (right panel) as functions of \(x=-\log(-k\tau)\). The vertical dotted line at \(x=0\) denotes horizon crossing. The second squeezing channel begins to grow earlier than the first, while both amplitudes develop pronounced late-time fluctuations. The squeezing angles approach different superhorizon values, reflecting the distinct phase evolution of the two channels.}
    \label{fig:squeezing paramaters}
\end{figure}

Fig. \ref{fig:squeezing paramaters} shows that the two squeezing channels remain weak in the subhorizon regime and separate after horizon crossing. The second channel develops significant squeezing earlier, whereas the first channel shows a delayed but rapid growth. At late times, both \(r_{1k}\) and \(r_{2k}\) exhibit fluctuations around an overall increasing background. This behavior follows from the different Hamiltonian coefficients \(A_{11}\) and \(A_{22}\), which are determined by different combinations of the coupled curvature and entropy mode functions. The squeezing angles show a different behavior. After an initial phase evolution, \(\varphi_{1k}\) and \(\varphi_{2k}\) approach distinct superhorizon values. Since the instantaneous growth rate of $r_{ik}$ is proportional to $\sin(2\varphi_{ik}-\arg A_{ii})$, the relative phase directly controls whether squeezing is instantaneously enhanced or suppressed. The different phase histories of the two squeezing eigenchannels therefore provide a natural origin for their different growth times and subsequent oscillatory structures. Inflationary particle production is naturally described as a squeezing process \cite{Albrecht:1992kf}. Previous studies have also shown that the evolution of quantum complexity in inflation is strongly correlated with the squeezing amplitude and squeezing phase, and can exhibit oscillatory behavior for modified dispersion relations and non-trivial sound speed \cite{Li:2024kfm,Li:2023ekd,Liu:2021nzx}. In the present two-field system, the same squeezing mechanism is enlarged to a four-mode structure with additional inter-field correlations \cite{Colas:2021llj,Bolis:2018jmo}.

\section{Krylov complexity of two-field inflation}
\label{sec:Krylov complexity of two-field inflation}

Krylov complexity provides a useful measure of operator growth and quantum information spreading \cite{Parker:2018yvk}. It has recently been applied to inflationary cosmology \cite{Li:2024kfm}. Further studies considered modified dispersion relations \cite{Li:2023ekd}, thermal states in the early Universe \cite{Li:2024ljz}, and modified-gravity inflation \cite{Li:2026jxx}. These works mainly focus on single-field or effectively single-field systems. In two-field inflation, the coupling between curvature and entropy perturbations introduces mode mixing and additional squeezing channels. These effects may provide new channels for quantum information spreading. Therefore, we extend Krylov complexity to two-field inflation to study the effects of multi-field interactions on primordial quantum information.

The growth of a quantum operator can be described in Krylov space. We first introduce the Liouvillian operator
\begin{equation}
    \mathcal{L}_\mathcal{O}=[H,\mathcal{O}],
\end{equation}
where \(H\) is the Hamiltonian and \(\mathcal{O}\) is the initial operator. The operator space is equipped with the inner product
\begin{equation}
    (\mathcal{O}_1|\mathcal{O}_2)
=
\frac{{\rm Tr}(\mathcal{O}_1^\dagger\mathcal{O}_2)}
{{\rm Tr}(1)}.
\end{equation}
Starting from the normalized operator \(|\mathcal{O}_0)\), an orthonormal Krylov basis is constructed by the Lanczos algorithm \cite{Parker:2018yvk},
\begin{equation}
    |A_n)=\mathcal{L}|\mathcal{O}_{n-1})-b_{n-1}|\mathcal{O}_{n-2}),\qquad b_n=\sqrt{(A_n|A_n)},
\qquad
|\mathcal{O}_n)=\frac{|A_n)}{b_n}.
\end{equation}
Therefore, the action of the Liouvillian becomes tridiagonal in the Krylov basis,
\begin{equation}
    \mathcal{L}|\mathcal{O}_n)=b_{n+1}|\mathcal{O}_{n+1})+b_n|\mathcal{O}_{n-1}),
\end{equation}
where \(b_n\) are the Lanczos coefficients. Thus, the original operator dynamics is mapped onto an effective one-dimensional semi-infinite chain, with \(b_n\) playing the role of hopping amplitudes. 
The time-dependent operator can then be expanded in the Krylov basis as
\begin{equation}
    |\mathcal{O}(t))=\sum_{n=0}^{\infty}i^n\phi_n(t)|\mathcal{O}_n),
\end{equation}
with the initial condition
\begin{equation}
    \phi_n(0)=\delta_{n0}.
\end{equation}
The Heisenberg equation gives the evolution equation for the Krylov wave function,
\begin{equation}
    \dot{\phi}_n(t)=b_n\phi_{n-1}(t)-b_{n+1}\phi_{n+1}(t),
\end{equation}
together with
\begin{equation}
    \sum_{n=0}^{\infty}|\phi_n(t)|^2=1.
\end{equation}
Hence, \(|\phi_n(t)|^2\) gives the probability distribution of the evolving operator along the Krylov chain. Its average position defines the operator Krylov complexity
\begin{equation}
    K(t)=\sum_{n=0}^{\infty}n\,|\phi_n(t)|^2,
\end{equation}
which is commonly referred to as the Krylov complexity. A larger \(K(t)\) means that the operator spreads toward higher Krylov levels and becomes more complex. For generic non-integrable systems, Parker et al. proposed the asymptotic behavior
\begin{equation}
    b_n\simeq \alpha n,
\end{equation}
which leads to exponential spreading in Krylov space,
\begin{equation}
    K(t)\sim e^{2\alpha t}.
\end{equation}
Thus, the Lanczos coefficients determine the growth of Krylov complexity and provide a direct connection between operator growth and quantum information spreading \cite{Parker:2018yvk}.

\subsection{Numerical behavior of the Krylov complexity}

For quantum states, an analogous state-space or spread-complexity construction characterizes the spreading of a time-evolving state over a Krylov basis \cite{Balasubramanian:2022tpr}. In the present squeezed-state setting, we use this state-space realization. The pair-occupation basis generated by the quadratic Hamiltonian forms a two-dimensional ladder labeled by ($n,m$). We therefore characterize the spreading of the four-mode squeezed state over this basis.

For the two-field system, the quantum state contains two squeezing eigenchannels in the diagonal squeezing basis. The four-mode squeezed state \eqref{eq:four-mode squeezed state} has been obtained in Sec. \ref{sec:Evolution Equations of the squeezing Parameters}. In the present representation, we identify the two squeezing ladders with the corresponding two-dimensional Krylov-lattice basis states. The state can therefore be expressed as
\begin{equation}
    |\Psi_k(\tau)\rangle
=
\sum_{n,m=0}^{\infty}
\phi_{n,m}(\tau)|K_{n,m}\rangle .
\end{equation}
By comparing the two expressions, the Krylov wave function is directly obtained as
\begin{equation}
    \phi_{n,m}(\tau)=\frac{e^{i\chi_k(\tau)}}{\cosh r_{1k}\cosh r_{2k}}(-1)^{n+m}e^{2in\phi_{1k}}e^{2im\phi_{2k}}\tanh^n(r_{1k})\tanh^m(r_{2k}).
\end{equation}
Its probability distribution is
\begin{equation}
    |\phi_{n,m}|^2=\frac{\tanh^{2n}(r_{1k})\tanh^{2m}(r_{2k})}{\cosh^2(r_{1k})\cosh^2(r_{2k})}.
\end{equation}
The overall phase \(\chi_k\) and the squeezing angles \(\varphi_{1k}\) and \(\varphi_{2k}\) disappear after taking the absolute square. The probability satisfies
\begin{equation}
    \sum_{n,m=0}^{\infty}|\phi_{n,m}|^2=1.
\end{equation}
For the two independent squeezing channels, we take the total Krylov level to be \(n+m\). The generalized state-space spread Krylov complexity is then defined as
\begin{equation}
    K_k=\sum_{n,m=0}^{\infty}(n+m)|\phi_{n,m}|^2 .
\end{equation}
Substituting the Krylov wave function gives
\begin{equation}
    K_k=\frac{1}{\cosh^2 r_{1k}\cosh^2 r_{2k}}\sum_{n,m=0}^{\infty}
(n+m)\tanh^{2n}r_{1k}\tanh^{2m}r_{2k}.
\end{equation}
Using the geometric-series relations
\begin{equation}
    \sum_{n=0}^{\infty}x^n=\frac{1}{1-x},\qquad\sum_{n=0}^{\infty}nx^n=\frac{x}{(1-x)^2},
\end{equation}
we finally obtain
\begin{equation}
    K_k=\sinh^2 r_{1k}+\sinh^2 r_{2k}
\end{equation}
Therefore, the total Krylov complexity can be separated into the contributions from the two squeezing channels,
\begin{equation}
    K_{1k}=\sinh^2 r_{1k},\qquad K_{2k}=\sinh^2 r_{2k},
\end{equation}
so that
\begin{equation}
    K_k=K_{1k}+K_{2k}.
\end{equation}
This result shows that the Krylov complexity is directly determined by the two squeezing strengths. The squeezing angles affect the phase of the Krylov wave function, but they do not contribute directly to \(K_k\). Thus, once \(r_{1k}\) and \(r_{2k}\) are obtained from the coupled evolution equations, the Krylov complexity can be calculated directly.

\begin{figure}[htbp]
    \centering
    \begin{minipage}{0.5\textwidth}
        \centering
        \includegraphics[width=\linewidth]{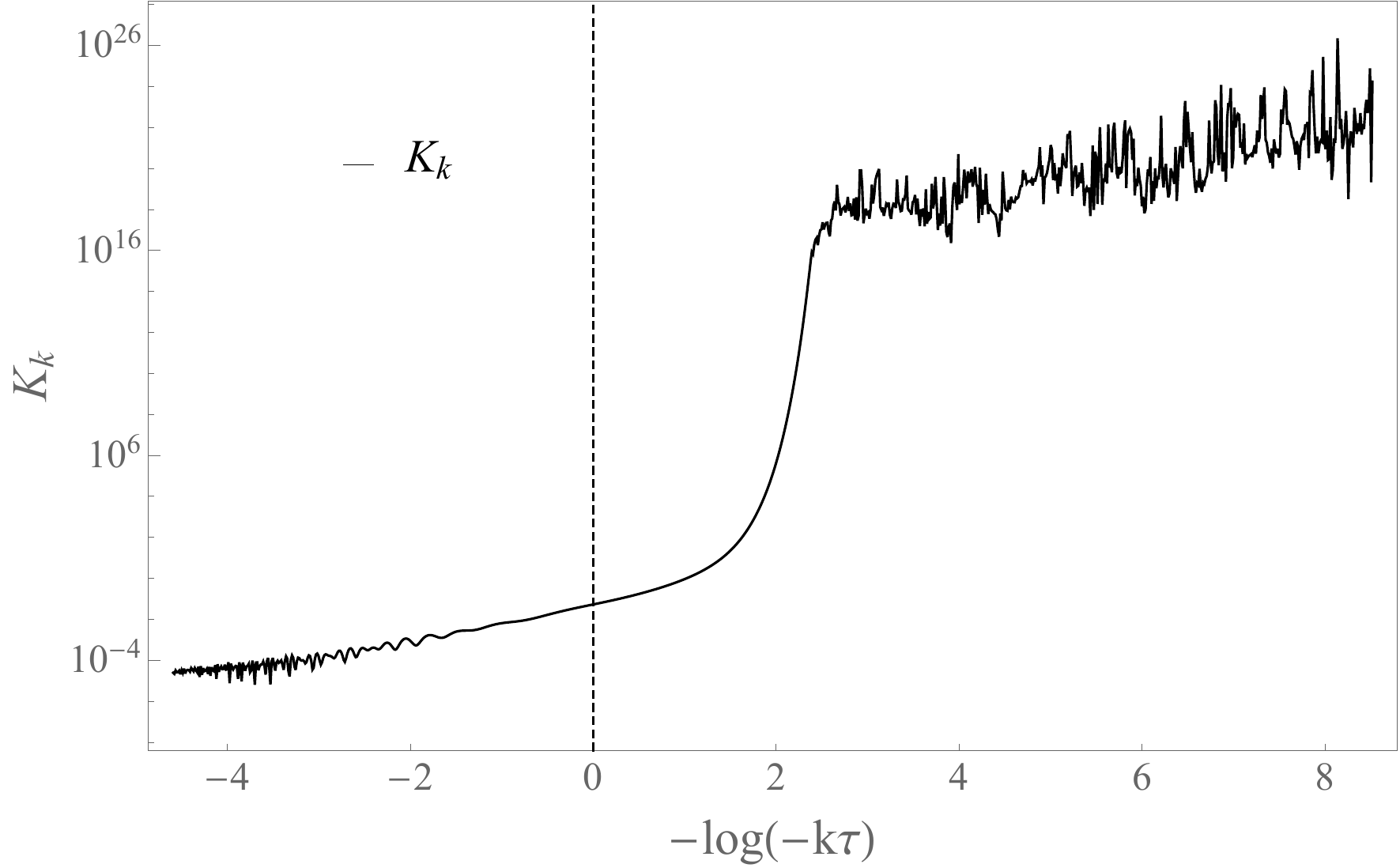}
    \end{minipage}

    \vspace{1em}
    
    \begin{minipage}{0.45\textwidth}
        \centering
        \includegraphics[width=\linewidth]{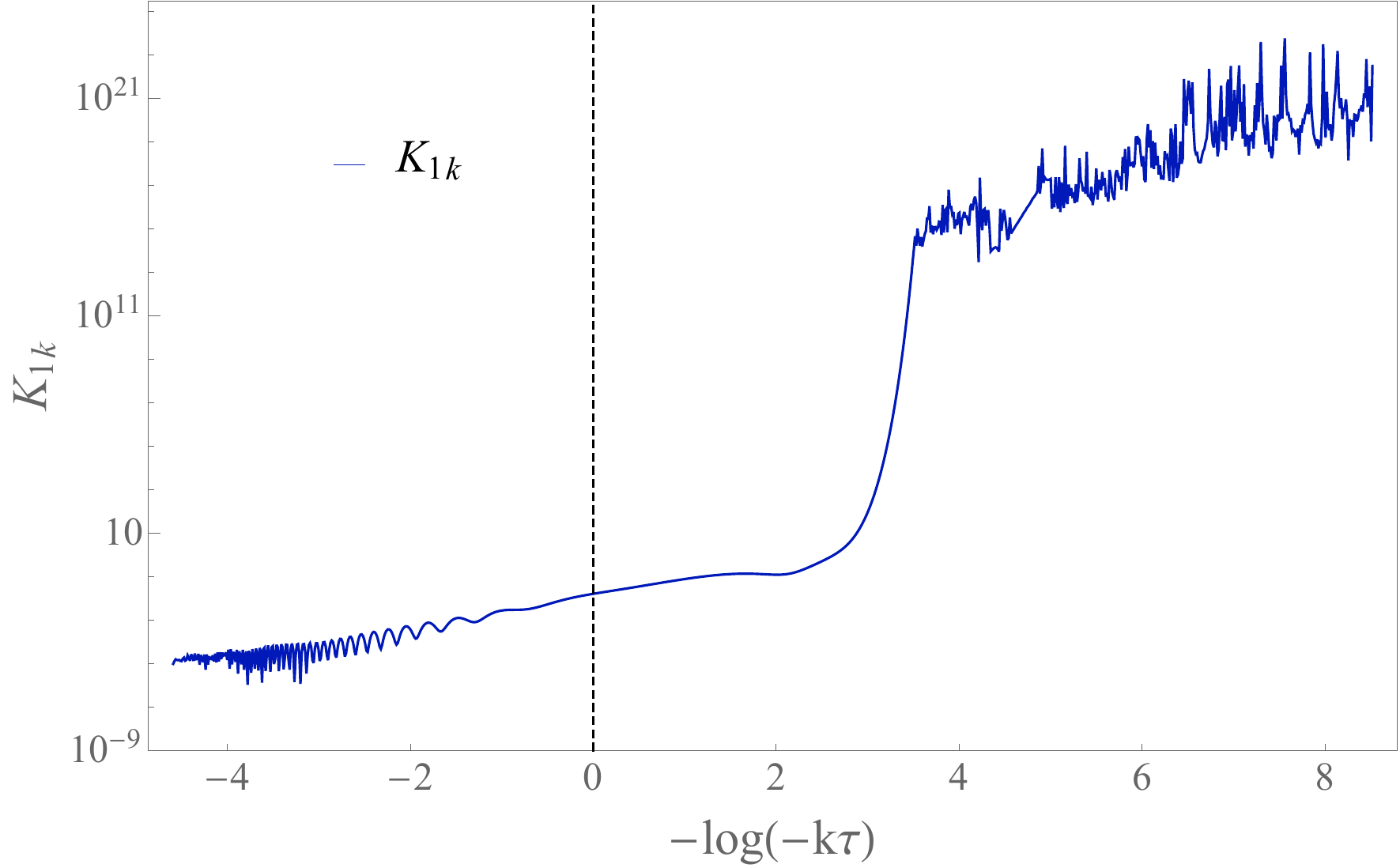}
    \end{minipage}
    \hfill
    \begin{minipage}{0.45\textwidth}
        \centering
         \includegraphics[width=\linewidth]{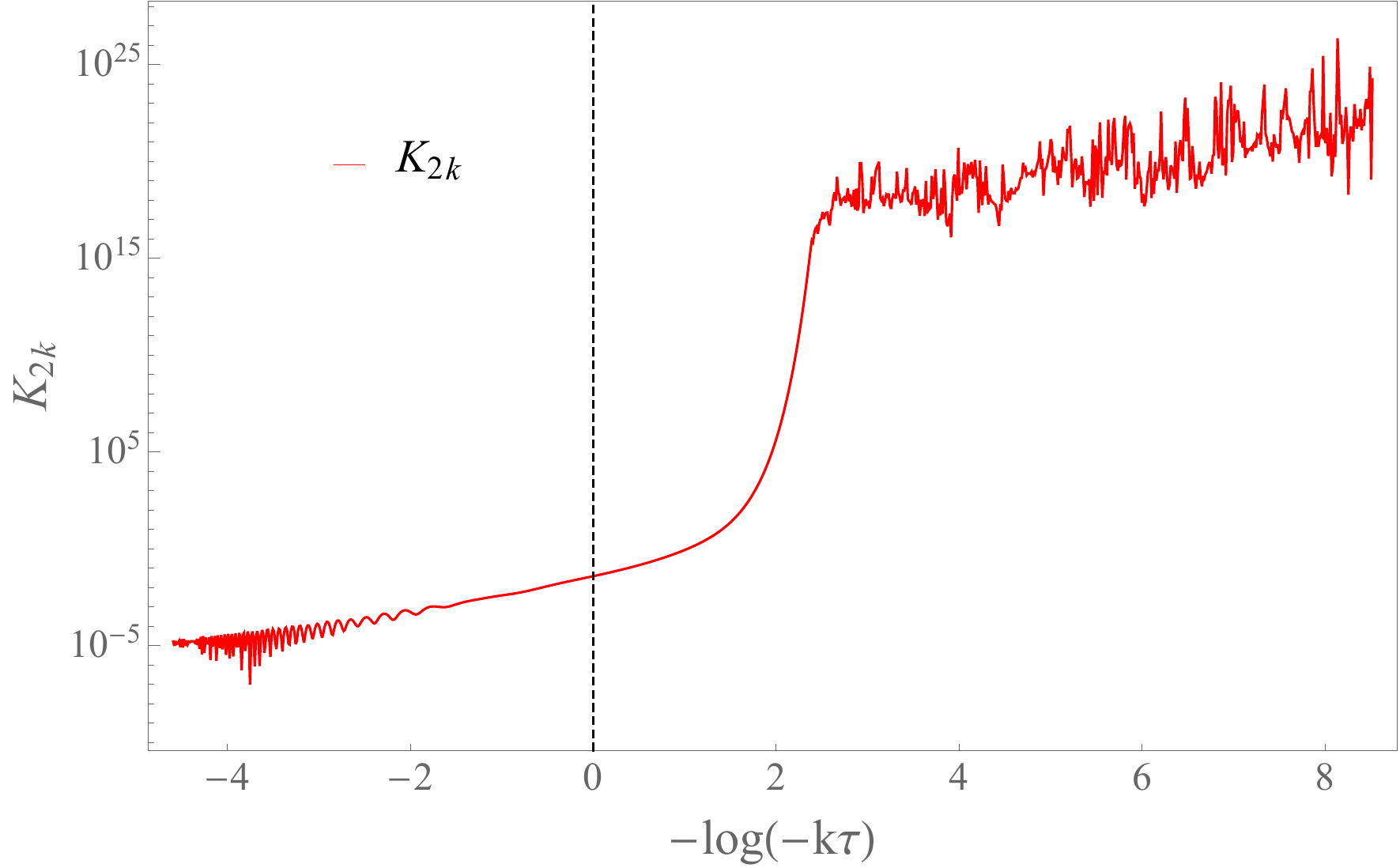}
    \end{minipage}
    \caption{shows the two channel contributions \(K_{1k}\) and \(K_{2k}\) and their sum \(K_k\). All three quantities remain small in the subhorizon regime. After horizon crossing, \(K_{2k}\) grows first and dominates the intermediate stage, reflecting the earlier growth of \(r_{2k}\). The first contribution \(K_{1k}\) rises later. At late times, both channels become large and inherit the fluctuations of the corresponding squeezing strength. Since \(K_{ik}=\sinh^2r_{ik}\), even moderate changes in \(r_{ik}\) produce large changes in the Krylov complexity in the strongly squeezing regime. The two contributions also exhibit distinct oscillatory structures, which originate from the phase-dependent squeezing dynamics and become strongly amplified in the large-squeezing regime.}
    \label{fig:Krylov complexity}
\end{figure}

Fig. \ref{fig:Krylov complexity} shows the evolution of the two channel contributions \(K_{1k}\) and \(K_{2k}\), together with the total Krylov complexity \(K_k\). Before horizon crossing, all three quantities remain small because both squeezing amplitudes are weak. After horizon exit, the two channels exhibit clearly different growth histories. The second channel grows earlier and dominates the intermediate stage, whereas the first channel shows a delayed but rapid enhancement. These channels should not be identified separately with the curvature and entropy fields, since each squeezing eigenchannels contains both curvature and entropy components. The difference between \(K_{1k}\) and \(K_{2k}\) is therefore a dynamical consequence of the coupled two-field evolution. The oscillatory behavior can be traced back to the phase-dependent squeezing source \(\sin(2\phi_{ik}-\arg A_{ii})\). Rapid phase evolution in the subhorizon regime produces small oscillatory modulations, while after horizon exit the growth of the squeezing amplitudes strongly amplifies the remaining phase modulation through \(K_{ik}=\sinh^2r_{ik}\). The oscillations of the two channels therefore have the same phase-sensitive origin but different amplitudes and onset times.

\subsection{Single-field limit and consistency check}

It is useful to verify that the present two-field construction reduces to the standard single-field result. The quantities \(\psi\), \(\mu\), \(\Omega\), and \(\lambda=2\Omega/H\) are associated with the additional entropy degree of freedom and its coupling to the curvature perturbation. In particular, the coupling enters through \(D_\tau\xi=\xi'-aH\lambda\psi/\sqrt{2\epsilon}\) and through the mixing term \(H_{\rm mix}\propto\lambda(\Pi_\xi\psi+\psi\Pi_\xi)\). Thus, \(\lambda\to0\) removes the direct curvature–entropy mixing, but this limit alone still describes two decoupled scalar sectors rather than a genuine single-field system.

The standard single-field limit is obtained by further restricting the theory to the curvature sector, \(\psi=\pi_\psi=0\). The entropy mass \(\mu\) then becomes irrelevant. Introducing the canonical Mukhanov variable \(v=z\xi\), with \(z=a\sqrt{2\epsilon}\), the remaining action takes the usual single-field form,
\[
S_{\rm sf}
=
\frac12\int d^3x\,d\tau
\left[
v'^2-(\nabla v)^2+\frac{z''}{z}v^2
\right].
\]
Therefore, the two-field system reduces to the standard canonical scalar perturbation. In the deep subhorizon regime, where \(k^2\gg |z''/z|\), the mode frequency approaches the comoving frequency \(k\). The diagonal Hamiltonian coefficient also simplifies in the single-field limit. After setting \(\lambda\to0\) and removing the entropy sector, the coefficient becomes $B_k=2a^2\epsilon\left(|\xi_k'|^2+k^2|\xi_k|^2\right)$. Introducing the canonical variable \(v_k=z\xi_k\), with \(z=a\sqrt{2\epsilon}\), this can be written as
\begin{equation}
    B_k=\left|v_k'-\frac{z'}{z}v_k\right|^2+k^2|v_k|^2.
\end{equation}
Thus, \(B_k\) is generally time dependent and is not exactly equal to the comoving wavenumber \(k\). For the Bunch–Davies mode in the deep subhorizon regime, \(|k\tau|\gg1\), the background-dependent corrections become negligible and \(B_k\) approaches the free-field value \(k\) with the normalization adopted here. Therefore, the present definition is consistent with the free-field behavior only asymptotically in the subhorizon limit.

The same reduction occurs in the squeezing and Krylov descriptions. Removing the entropy sector leaves only one two-mode squeezing channel, so the \(Sp(4,\mathbb R)\) four-mode structure reduces to the usual \(Sp(2,\mathbb R)\) two-mode system. Formally, \(r_{2k}\to0\), and the two-field Krylov complexity $K_k=\sinh^2r_{1k}+\sinh^2r_{2k}$ reduces to
\begin{equation}
    K_k^{\rm(sf)}=\sinh^2r_k.
\end{equation}
The two-dimensional Krylov basis labeled by \((n,m)\) simultaneously reduces to the standard one-dimensional Krylov chain. This provides a direct consistency check that the present two-field formulation contains the previously studied single-field inflationary Krylov dynamics as a limiting case.

\section{Krylov entropy and Lanczos coefficients}
\label{sec:Krylov entropy and Lanczos coefficients}

Using the Krylov probability distribution obtained in Sec. \ref{sec:Krylov complexity of two-field inflation},
\[
P_{n,m}
=
\frac{\tanh^{2n}r_{1k}\tanh^{2m}r_{2k}}
{\cosh^2r_{1k}\cosh^2r_{2k}},
\]
we now study the corresponding Krylov entropy and Lanczos coefficients.

\subsection{Krylov entropy}
The Krylov entropy measures the spreading of the state over the Krylov basis. Krylov entropy has also been used in recent early-Universe studies to characterize the broadening of the Krylov probability distribution and to distinguish different inflationary or thermal scenarios \cite{Li:2024ljz,Liu:2025caj,Zhai:2024odw}. It is defined as
\begin{equation}
    S_K=-\sum_{n,m=0}^{\infty}P_{n,m}\ln P_{n,m}.
\end{equation}
Using the factorized probability distribution, we obtain $S_K=S_{K,1}+S_{K,2}$
\begin{equation}
    \begin{split}
      &  S_{K,1}=\cosh^2r_{1k}\ln\!\left(\cosh^2r_{1k}\right)-\sinh^2r_{1k}\ln\!\left(\sinh^2r_{1k}\right),\\&
      S_{K,2}
=
\cosh^2r_{2k}
\ln\!\left(\cosh^2r_{2k}\right)
-
\sinh^2r_{2k}
\ln\!\left(\sinh^2r_{2k}\right).
    \end{split}
\end{equation}
Thus, the total Krylov entropy is
\begin{equation}
    \begin{split}
        S_K&=\cosh^2r_{1k}\ln\!\left(\cosh^2r_{1k}\right)-\sinh^2r_{1k}\ln\!\left(\sinh^2r_{1k}\right)
\\&+\cosh^2r_{2k}\ln\!\left(\cosh^2r_{2k}\right)-\sinh^2r_{2k}\ln\!\left(\sinh^2r_{2k}\right).
    \end{split}
\end{equation}
The result depends only on the two squeezing strengths. The squeezing angles do not enter because the phase factors disappear in \(P_{n,m}=|\varphi_{n,m}|^2\). The relation with the Krylov complexity is also simple. In the previous Sec. \ref{sec:Krylov complexity of two-field inflation}, we obtained
\[
K_{1k}=\sinh^2r_{1k},
\qquad
K_{2k}=\sinh^2r_{2k},\qquad K_k=K_{1k}+K_{2k}.
\]
Therefore,
\begin{equation}
    S_{K,i}=(1+K_{ik})\ln(1+K_{ik})-K_{ik}\ln K_{ik}
\end{equation}
for \(i=1,2\). The total entropy becomes
\begin{equation}
\begin{split}
      S_K&=(1+K_{1k})\ln(1+K_{1k})-K_{1k}\ln K_{1k}
\\&+(1+K_{2k})\ln(1+K_{2k})-K_{2k}\ln K_{2k}.
\end{split}
  \end{equation}
This relation directly connects the Krylov entropy with the Krylov complexity of the two squeezing channels.

\begin{figure}
    \centering
    \includegraphics[width=0.75\linewidth]{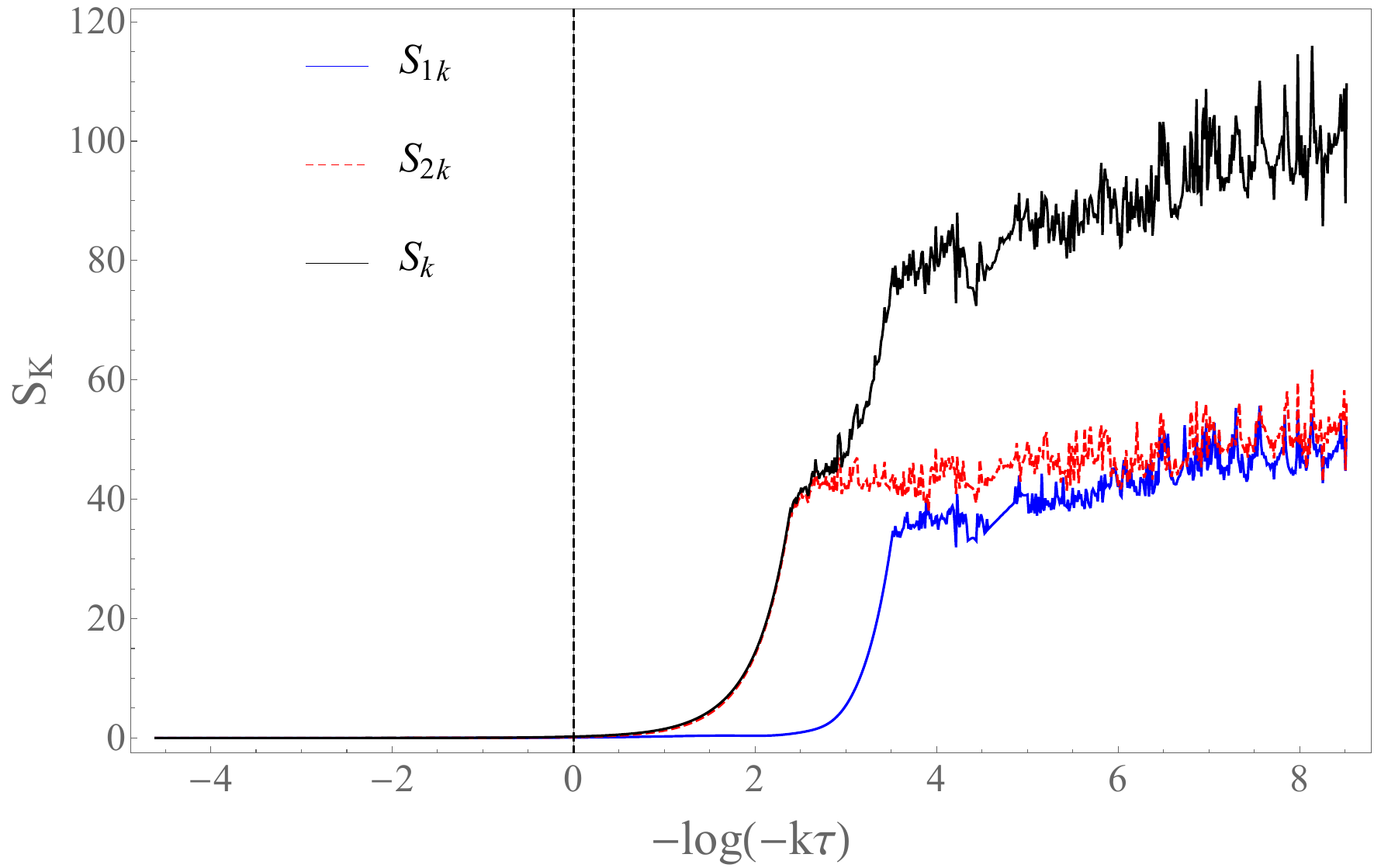}
    \caption{Evolution of the Krylov entropy during two-field inflation. The blue solid and red dashed curves denote the two channel contributions \(S_{1k}\) and \(S_{2k}\), respectively, while the black solid curve represents the total entropy \(S_k=S_{1k}+S_{2k}\). The vertical dashed line marks horizon crossing. The entropy grows after horizon exit as squeezing develops, with the two channels showing different onset times and late-time oscillatory behavior.}
    \label{fig:Krylov entropy}
\end{figure}
Fig. \ref{fig:Krylov entropy} shows the Krylov entropy of the two squeezing channels. Before horizon crossing, both contributions remain small because the Krylov probability distribution is concentrated near the initial state. After horizon exit, \(S_{2k}\) grows earlier, followed by a delayed increase of \(S_{1k}\). The total entropy therefore exhibits a two-stage growth pattern. At late times, the entropy curves inherit the fluctuations of the squeezing strength. In contrast to the exponentially sensitive behavior of \(K_{ik}=\sinh^2r_{ik}\), the entropy grows more slowly in the strongly squeezing regime, where \(S_{ik}\simeq \ln K_{ik}+1\simeq2r_{ik}+1-\ln4\). Thus, the Krylov complexity measures the mean extent of spreading in the Krylov basis, whereas the Krylov entropy characterizes the broadening of the associated probability distribution.

\subsection{Lanczos coefficients}

The same four-mode basis can also be used to describe the transitions in Krylov space. We define
\begin{equation}
    |K_{n,m}\rangle=|n_{1,k},n_{1,-k},m_{2,k},m_{2,-k}\rangle .
\end{equation}
For the first channel, the pair operators satisfy
\begin{equation}
    \begin{split}
        \hat a_1(k)\hat a_1(-k)|K_{n,m}\rangle=n|K_{n-1,m}\rangle ,\qquad  \hat a_1^\dagger(k)
\hat a_1^\dagger(-k)|K_{n,m}\rangle=(n+1)|K_{n+1,m}\rangle .
    \end{split}
\end{equation}
For the second channel,
\begin{equation}
    \begin{split}
        \hat a_2(k)\hat a_2(-k)|K_{n,m}\rangle=m|K_{n,m-1}\rangle ,\qquad \hat a_2^\dagger(k)
\hat a_2^\dagger(-k)
|K_{n,m}\rangle
=
(m+1)|K_{n,m+1}\rangle .
    \end{split}
\end{equation}
In the diagonal squeezing basis, the pair-production coefficients \(A_{11}\) and \(A_{22}\) generate transitions along the two independent directions of the two-dimensional Krylov lattice. The corresponding quantities
\begin{equation}
    b_n^{(1)}(\tau)=\frac{|A_{11}(\tau)|}{2}n,
\qquad
b_m^{(2)}(\tau)=\frac{|A_{22}(\tau)|}{2}m,
\end{equation}
play the role of channel-resolved Lanczos coefficients for the two squeezing directions. They should be distinguished from the Lanczos coefficients of a single one-dimensional Krylov chain obtained by applying the Lanczos recursion directly to the full two-field Hamiltonian. In the present representation, \(b_n^{(1)}\) and \(b_m^{(2)}\) characterize the hopping amplitudes along the two independent Krylov directions. The number-conserving coefficients contribute to the diagonal part of the Krylov evolution,
\begin{equation}
    a_{n,m}=(2n+1)B_{11}+(2m+1)B_{22}.
\end{equation}
Thus, the two-field Krylov dynamics has a two-dimensional structure. The states are labeled by \((n,m)\), and the two directions are controlled by different Hamiltonian coefficients.

\begin{figure}
    \centering
    \includegraphics[width=0.75\linewidth]{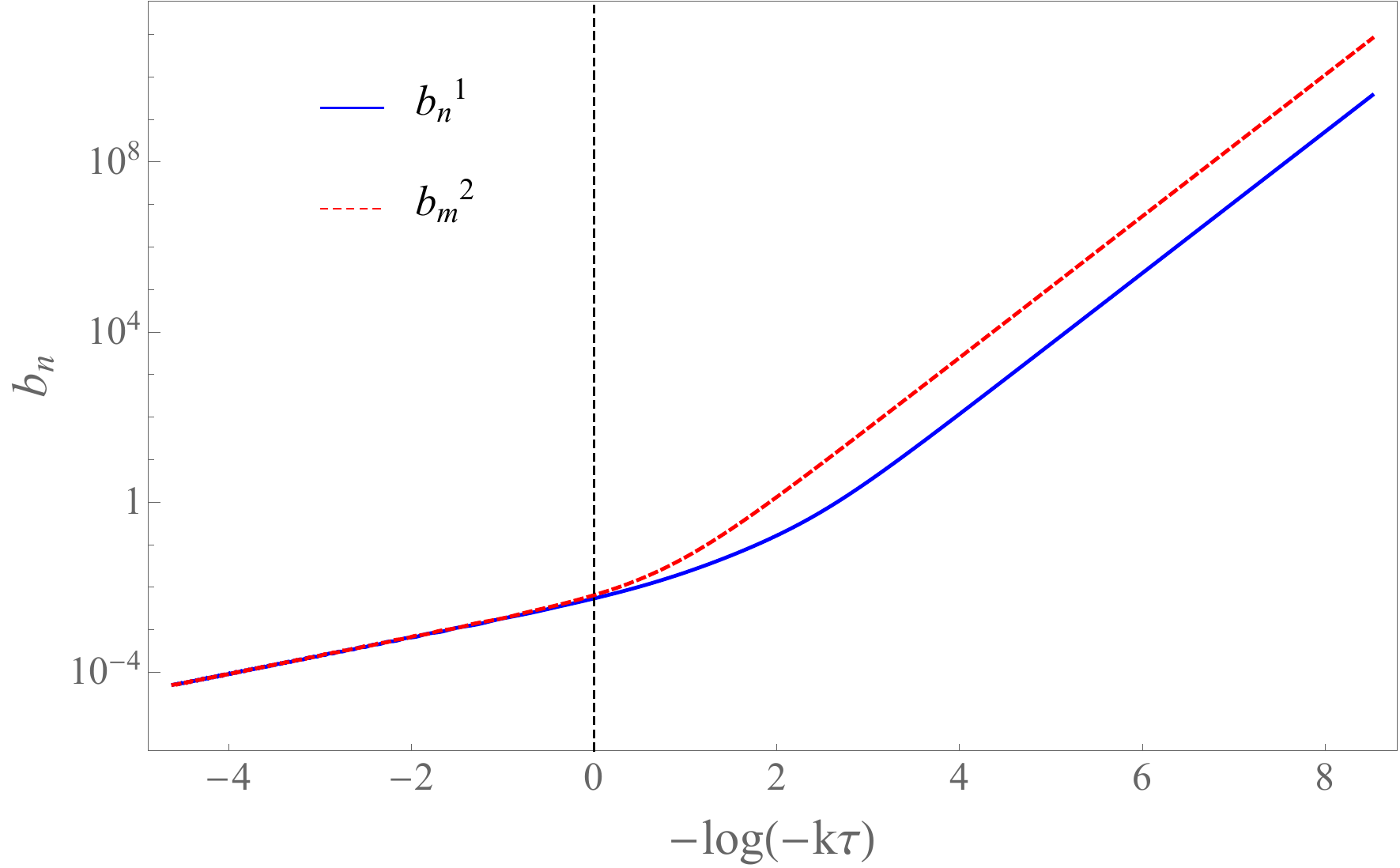}
    \caption{Evolution of the channel-resolved Lanczos coefficients \(b_n^{(1)}\) and \(b_m^{(2)}\) for \(n=m=1\) during two-field inflation. The blue solid and red dashed curves correspond to the first and second perturbation channels, respectively. The vertical dashed line denotes horizon crossing at \(-\log(-k\tau)=0\). Both coefficients increase during the superhorizon evolution, while the second channel develops a larger hopping coefficient at late times.}
    \label{fig:bn}
\end{figure}

Fig. \ref{fig:bn} shows the time evolution of the channel-resolved Lanczos coefficients for \(n=m=1\). Both coefficients are small at early times and increase strongly during the superhorizon evolution. The second channel becomes larger than the first, consistent with the difference between \(|A_{22}|\) and \(|A_{11}|\). The increase of these coefficients indicates stronger transitions between neighboring occupation sectors of the two-dimensional Krylov lattice. This hierarchy is consistent with the earlier onset of squeezing and Krylov growth in the second channel.

Taken together, Figs. \ref{fig:squeezing paramaters}--\ref{fig:bn} reveal a consistent two-channel picture of Krylov-space evolution. The second channel responds earlier to the curvature–entropy interaction and dominates the intermediate stage, whereas the first channel exhibits a delayed enhancement. The phase-dependent squeezing dynamics generates oscillatory behavior in both channels, which is strongly amplified in the Krylov complexity once large squeezing is reached. The same channel hierarchy is reflected in the Krylov entropy and in the channel-resolved Lanczos coefficients.

\section{Conclusions and Outlook}
\label{sec:Conclusions and Outlook}

The two squeezing eigenchannels show distinct dynamical responses to the underlying curvature–entropy coupling. The second channel responds earlier and dominates the intermediate stage, whereas the first exhibits a delayed enhancement. Their phase-sensitive squeezing dynamics generates different oscillatory Krylov patterns across horizon crossing. The nonlinear relation \(K_{ik}=\sinh^2r_{ik}\) strongly amplifies these modulations in the large-squeezing regime.
\[
K_k=K_{1k}+K_{2k}
=\sinh^2r_{1k}+\sinh^2r_{2k}.
\]
The second channel dominates the intermediate stage because its squeezing develops earlier. The first channel becomes important at later times. The total Krylov complexity therefore records the different squeezing histories of the coupled perturbation modes. An important feature of the present two-field dynamics is the oscillatory structure of the Krylov complexity around and after horizon crossing. These oscillations originate from the phase-sensitive squeezing source \(\sin(2\varphi_{ik}-\arg A_{ii})\). In the subhorizon regime, the rapid evolution of the relative phase produces small modulations of the squeezing rates. After horizon exit, the squeezing amplitudes become large and the remaining phase modulation is strongly amplified through the nonlinear relation \(K_{ik}=\sinh^2 r_{ik}\). The two channels therefore exhibit different oscillatory patterns because they are controlled by different Hamiltonian coefficients constructed from the coupled curvature and entropy modes. This channel dependence is absent in the usual single-channel squeezed-state description and represents the main qualitative difference introduced by the two-field dynamics.

We also evaluated the Krylov entropy. Because the probability distribution factorizes between the two squeezing channels, the total entropy satisfies
\[
S_K=S_{K,1}+S_{K,2}.
\]
The Krylov entropy provides a complementary measure of the broadening of the same Krylov probability distribution and inherits the two-stage, channel-dependent evolution.

Finally, we derived channel-resolved Lanczos coefficients for the two-dimensional occupation lattice. Their growth during the superhorizon evolution indicates an enhancement of transitions between neighboring Krylov sectors. The two hopping directions evolve differently because they are controlled by different Hamiltonian coefficients. This provides another manifestation of the two-channel structure generated by curvature–entropy coupling.

These results extend the squeezing-state Krylov description from an effectively single-field system to the simplest coupled multi-field setting. A natural next step is to apply the construction to three-field and general \(N\)-field inflation. In such systems, several entropy directions and mixing channels are present, and the Gaussian state is associated with a larger \(Sp(2N,\mathbb{R})\) structure. It would also be interesting to investigate reduced-state dynamics and environmental interactions, where decoherence and information loss can be defined explicitly rather than inferred from unitary Krylov-space spreading.

\acknowledgments
TL is supported by the Guizhou Provincial Natural Science Foundation (Grant No.: 2025YJSKYJJ178) and the Natural Science Foundation of Guizhou Minzu University (Grant No.: GZMUBS7). LH is funded by the NSFC grant NO. 12665009 and Hunan Provincial Department of Education Project (Grant No.25B0480). HB is supported by the National Natural Science Foundation of China under Grant No.12265010, the Project of Guizhou Provincial Department of Science and Technology under Grants No.MS[2025]219 and No.CXTD[2025]030.

\bibliography{refs.bib}
\end{document}